\documentclass[11pt,aps,prd,onecolum,amsmath,amssymb,superscriptaddress,floatfix]{revtex4-2}

\usepackage{graphicx}
\usepackage{dcolumn}
\usepackage{bm}
\usepackage[caption=false]{subfig}
\usepackage{booktabs}
\usepackage{xcolor}
\usepackage{tikz}
\usepackage{overpic}
\usepackage[colorlinks=true,linkcolor=blue,citecolor=blue,urlcolor=blue]{hyperref}

\graphicspath{{Figures/}}

\newif\ifauthornotes
\authornotesfalse

\newcommand{\figorplaceholder}[2]{%
  \IfFileExists{Figures/#2}{\includegraphics[width=#1]{#2}}%
  {\figplaceholder{#1}{2.8cm}{add \texttt{\detokenize{Figures/#2}}\\(from ~\cite{Singh:2025ogp})}}}
\newcommand{\figplaceholder}[3]{%
  \fbox{\parbox[c][#2][c]{#1}{\centering\footnotesize
  \textbf{Figure placeholder}\\[3pt]#3}}}

\newcommand{\Op}{\mathcal{O}}
\newcommand{\vev}[1]{\langle #1\rangle}
\newcommand{\muB}{\mu_B}

\begin{document}

\title{Competition and coexistence of baryonic condensates in AdS hardwall model}

\author{Akash Singh}
\email{akash.singh@wits.ac.za}
\affiliation{Mandelstam Institute for Theoretical Physics, School of Physics, and NITheCS,
University of the Witwatersrand, Johannesburg, WITS 2050, South Africa}

\date{\today}

\begin{abstract}
Cold dense QCD admits several order parameters for baryon-number breaking, among them diquark and dibaryon condensates. We ask whether they exclude one another or coexist, using a five-dimensional AdS hardwall model with full backreaction in which two charged scalars, dual to operators of charges $q_i$ and dimensions $\Delta_i$, couple to a common $U(1)_B$ gauge field.  With the diquark-like operator fixed at $q_1=2$, $\Delta_1=3$, both operators condense only within a narrow window of $\Delta_2$, which widens with $q_2$ and closes when the operators coincide. Outside the window, only a single condensate forms. The structure persists in both confined and deconfined phases. In the $\mu_B$--$T$ plane,
confined matter with both condensates is preferred at low temperature, below a
narrow band of deconfined matter with one condensate.
\end{abstract}

\maketitle

\section{Introduction}\label{sec:intro}

The phase structure of QCD at large baryon chemical potential and low temperature is still not fully understood~\cite{Brambilla:2014jmp, Fukushima:2010bq}. Lattice simulations, which settle most questions at $\muB=0$, are obstructed by the sign problem once $\muB\neq0$, and perturbative QCD becomes reliable only at baryon densities of order $\gtrsim40\,n_0$~\cite{Kurkela:2009gj}, where $n_0\simeq0.16~\text{fm}^{-3}$ is the nuclear saturation density, far above the few times $n_0$ reached in the cores of neutron stars \cite{Annala:2019puf}. The intermediate regime, which is the one realized in neutron star cores, therefore has to be explored with effective models and other nonperturbative methods~\cite{Haensel:2007yy, Schmitt:2010pn, Baym:2017whm}.

Part of the difficulty is that this regime offers many candidate ground states. As the density increases, nucleons are expected to pair into a neutron superfluid and a proton superconductor~\cite{ Sedrakian:2018ydt}, mesons may condense~\cite{Pethick:2015jma}, and at still higher densities quarks near the Fermi surface pair into color superconducting phases, with the color-flavor locked (CFL) phase as the ground state at asymptotically high density~\cite{Alford:1997zt, Alford:1998mk, Alford:2007xm}. Each of them would leave an imprint on the equation of state and hence on compact stars~\cite{Alford:2017qgh,Baym:2017whm}. Several of these phases break the same symmetry, $U(1)_B$, but through different operators. A quark bilinear carries quark number $2$, while the order parameter of a baryon-pair superfluid carries quark number $2 N_c$ (for $N_c=3$, it contains six quark fields). The two are not unrelated: in the CFL phase the gauge-invariant order parameter for $U(1)_B$ breaking is itself a six-quark operator with the quantum numbers of the $H$ dibaryon~\cite{Jaffe:1976yi,Alford:2007xm}, which is one of the observations behind the conjectured continuity between hyperonic and quark matter~\cite{Schafer:1998ef}. It is then natural to ask what happens when operators with different baryon charges and scaling dimensions become unstable at comparable densities. Does the first operator to condense suppress the others, or can several condensates coexist? Effective models such as the Nambu--Jona-Lasinio (NJL) model address this kind of interplay~\cite{Asakawa:1989bq,Buballa:2003qv}, and the competition between the chiral and diquark condensates in particular has been studied in detail~\cite{Berges:1998rc,Yamamoto:2007ah}. The answers depend on couplings that are poorly constrained at these densities, and model equations of state need not respect causality or thermodynamic stability once they are extrapolated beyond the region in which they were fitted~\cite{Baym:2017whm,Lindblom:2018rfr}.

Gauge/gravity duality~\cite{Maldacena:1997re, Gubser:1998bc, Witten:1998qj} offers a different handle on strongly coupled matter at finite density. Bottom-up models break conformal invariance with an infrared cutoff on the radial coordinate, the hardwall~\cite{Erlich:2005qh,DaRold:2005mxj,Herzog:2006ra}, or with a dilaton profile~\cite{Karch:2006pv, Jarvinen:2011qe}, while top-down constructions based on intersecting branes~\cite{Karch:2002sh,Sakai:2004cn,Sakai:2005yt} are more rigid but closer to string theory. Both have been used to construct equations of state for dense QCD matter and to compare them with neutron star observations~\cite{Hoyos:2016zke,Jokela:2018ers,BitaghsirFadafan:2020otb,Hoyos:2021uff,Jarvinen:2021jbd,Demircik:2021zll,Kovensky:2021kzl,Bartolini:2022rkl,Bartolini:2023wis,Jarvinen:2023jbr}. In this framework a condensate is described by a charged bulk scalar which develops a normalizable profile above a critical chemical potential. This is the mechanism behind holographic superconductors~\cite{Gubser:2008px,Hartnoll:2008vx,Hartnoll:2008kx,Hartnoll:2009sz,Herzog:2009xv,Cai:2015cya}. Holographic models of color superconductivity built on this mechanism~\cite{Basu:2011yg,BitaghsirFadafan:2018iqr,Ghoroku:2019trx,Faedo:2018fjw,Preau:2025ubr} have so far concentrated on a single condensing operator.

The competition between several orders has been studied mostly for holographic superconductors in black hole backgrounds. Basu \textit{et al.}~\cite{Basu:2010fa} considered two charged scalars coupled to the same gauge field in the probe limit and found that the condensation of one generally hinders the other. Musso~\cite{Musso:2013ija} showed, in a model with two gauge fields, that an attractive direct interaction between the scalars leads to competition and a repulsive one to enhancement. Cai \textit{et al.}~\cite{Cai:2013wma} included the backreaction in a two-scalar model without direct coupling and found five superconducting phases, three of which contain regions where both orders coexist. Related work has examined multi-band models with other symmetry structures~\cite{Huang:2011ac,Krikun:2013hwi}, the competition between $s$- and $p$-wave~\cite{Nie:2013sda,Kiritsis:2015hoa} and $s$- and $d$-wave~\cite{Li:2014wca} orders, and the interior of multi-band black holes~\cite{Zhang:2025hkb}; see~\cite{Cai:2015cya} for a review. A recurring outcome is that coexistence does occur, but only in narrow regions of parameter space~\cite{Basu:2010fa, Li:2014wca}. In the AdS soliton~\cite{Witten:1998zw, Horowitz:1998ha}, a holographic dual of a confining phase, single condensates were studied in~\cite{Nishioka:2009zj, Horowitz:2010jq}, and coexisting condensates have been found at zero temperature~\cite{Ran:2016cfh}. However, a careful study of coexisting condensates in confined phases at finite density in AdS/QCD-type models, or for the pair of operators relevant to dense QCD, is still needed.

In earlier work~\cite{Singh:2022obu,Singh:2024amm,Singh:2025ogp} we studied hardwall models of QCD, in particular a five-dimensional model at finite density in which the confined phase is described by a charged AdS (CAdS) geometry, with the charge carried by sources behind the infrared cutoff. Adding a charged scalar, we found that diquark-like ($q=2$, $\Delta=3$) and dibaryon-like ($q=2N_c$) condensates form in the confined phase at low temperature, and that dibaryon-like operators of dimension away from perturbative value condense before the diquark-like one~\cite{Singh:2025ogp}. 

In this paper, we study coexistence and competition of two condensates systematically. We introduce two charged scalars, dual to operators $\Op_1$ and $\Op_2$ with independent charges and dimensions. We work in both the confined phase, with the simplest infrared condition on the gauge field, and in the deconfined, charged black hole (CBH) phase. Our main results are the following. Keeping $\Op_1$ fixed as a diquark-like operator ($q_1=2$, $\Delta_1=3$), the two operators coexist only within a narrow window of $\Delta_2$, which we determine for $q_2=3,4$ and $6$. Outside the window the pressure is maximized by a single condensate. The window moves to larger $\Delta_2$ and widens as $q_2$ increases, and it closes at the degenerate point $(q_2,\Delta_2)=(2,3)$. At both edges of the window, the vanishing condensate goes to zero continuously.
As functions of the chemical potential, both condensates appear through
continuous transitions, and once present they persist. The sequence in which the operators condense depends on their charges and scaling dimensions.
The phase diagrams
in the $\muB$--$T$ plane have the same structure for the two cases we study, $q_2=3$ and the dibaryon-like $q_2=6$. At low
temperature, confined matter with both condensates is preferred throughout the
range of $\muB$ we computed.

The paper is organized as follows. In Sec.~\ref{sec:model} we set up the model and discuss the interpretation of the charges and dimensions. Sec.~\ref{sec:background} collects the solutions without condensates, the boundary conditions and physical scales, and the single-condensate results. Sec.~\ref{sec:cads} presents the two-field solutions in the confined phase and the coexistence windows. Sec.~\ref{sec:thermo} follows the confined phase as a function of the chemical potential, Sec.~\ref{sec:cbh} discusses the deconfined phase, and Sec.~\ref{sec:pd} assembles the phase diagrams. We conclude in Sec.~\ref{sec:discussion}. The appendices contain the holographic renormalization and a study of the dependence on the ultraviolet cutoff.

\section{Holographic model}\label{sec:model}

We consider Einstein--Maxwell gravity  coupled to two complex scalars, both minimally coupled to a single $U(1)$ gauge field $A_\mu$ dual to the baryon current,
\begin{multline}\label{eq:action}
S=\int d^5x\sqrt{-g}\,\bigg[\frac{1}{2\kappa^2}\left(R-2\Lambda\right)-\frac{1}{4g_5^2}F_{\mu\nu}F^{\mu\nu}
-\sum_{i=1}^{2}\lambda_i\left(|D_\mu\psi_i|^2+m_i^2|\psi_i|^2\right)\bigg],
\end{multline}
with $D_\mu\psi_i=(\partial_\mu-iq_iA_\mu)\psi_i$ and $2\Lambda=-12/L^2$ such that the geometry is asymptotically AdS$_5$. The couplings are,
\begin{equation}\label{eq:couplings}
\kappa^2=\frac{4\pi^2L^3}{N_c^2},\qquad g_5^2=\frac{24\pi^2L}{\theta N_fN_c}, \qquad \lambda_i=1,
\end{equation}
where $\theta$ removes the model dependence of the current--current correlator \cite{Singh:2024amm}. Since the scalar action is quadratic, each $\lambda_i$ can be absorbed into the normalization of $\psi_i$, and we set $\lambda_i=1$~\cite{Singh:2025ogp}. The backreaction of the gauge field on the geometry is controlled by
\begin{equation}\label{eq:zeta}
\zeta=\frac{2\kappa^2}{3g_5^2}=\frac{\theta N_fL^2}{9N_c},
\end{equation}
with the probe limit at $\zeta\to0$. We fix $\zeta=0.77$, the value that reproduces Boltzmann pressure of nuclear matter at high temperature~\cite{Hoyos:2016zke}. The dependence of the single-condensate results on $\zeta$ was studied in~\cite{Singh:2025ogp}. As in our earlier work, the use of holography involves an extrapolation from large $N_c$ to $N_c=3$ which is not justified from first principles, and our results should be read as qualitative statements about the structure of the competition rather than quantitative predictions. We work in a two-flavor system, $N_f=2$.

The charge $q_i$ and mass $m_i$ of $\psi_i$ fix the baryon charge and scaling dimension of $\Op_i$. We do not include a direct coupling between the scalars, such as a term $|\psi_1|^2|\psi_2|^2$, so the two condensates interact only through the gauge field and the metric. 

\subsection{Ansatz and equations of motion}

We take a static, homogeneous and isotropic configuration,
\begin{gather}
ds^2=-\frac{g(z)}{h(z)}L^2dt^2+\frac{L^2}{z^2}d\vec x^{\,2}+\frac{L^2}{g(z)}dz^2,\nonumber\\
A=\phi(z)\,dt,\qquad \psi_i=\psi_i(z),\label{eq:ansatz}
\end{gather}
where $z$ is the radial coordinate with the UV boundary at $z\to0$. In the gauge $A_z=0$ the radial component of Maxwell's equations, together with the scalar equations, forces the phases of $\psi_1$ and $\psi_2$ to be constant~\cite{Cai:2013wma}. One of them is removed by the gauge symmetry. The other, a constant relative phase, drops out of all equations because the scalars have no direct coupling. We may therefore take both scalars real without loss of generality.

We rescale coordinates and fields by $L$ so that they are dimensionless, and set $L=1$ in the numerical analysis. The equations of motion are

\begin{align}
g'-g\left(\frac{4}{z}+\frac{h'}{2h}\right)+4z&=\frac{2\zeta z}{L^2}\left[\frac{h\phi'^2}{4}+\sum_{i}\frac{g_5^2m_i^2 L\psi_i^2}{2}\right],\label{eq:EOMg}\\
\frac{h'}{h}-\frac{4}{z}&=\frac{2\zeta g_5^2z}{L^3}\sum_i\left(\psi_i'^2+\frac{q_i^2\psi_i^2\phi^2h}{g^2}\right),\label{eq:EOMh}\\
\phi''-\phi'\left(\frac{3}{z}-\frac{h'}{2h}\right)&=\frac{2g_5^2}{g L}\sum_iq_i^2\psi_i^2\phi,\label{eq:EOMphi}\\
\psi_i''+\psi_i'\left(\frac{g'}{g}-\frac{h'}{2h}-\frac{3}{z}\right)&=m_{i,\text{eff}}^2\,\psi_i,\label{eq:EOMpsi}
\end{align}
with $' \equiv \frac{d}{dz}$
and the effective mass
\begin{equation}\label{eq:meff}
m_{i,\text{eff}}^2=\frac{m_i^2}{g}-\frac{q_i^2h\phi^2}{g^2}.
\end{equation}
The two scalars appear in each other's equations only through $\phi$ in Eq.~\eqref{eq:EOMphi} and through $g$ and $h$ in Eqs.~\eqref{eq:EOMg} and \eqref{eq:EOMh}. Condensation of $\psi_i$ requires $m_{i,\text{eff}}^2$ to be sufficiently negative over an extended range of $z$~\cite{Gubser:2008px}. The chemical potential drives it negative through the second term of Eq.~\eqref{eq:meff}. 

Near the UV boundary,
\begin{gather}
\phi\simeq\mu-Qz^2,\qquad \psi_i\simeq\psi_i^{(s)}z^{\Delta_{i-}}+\psi_i^{(v)}z^{\Delta_{i+}},\nonumber\\
g\simeq z^2(1+cz^4),\qquad h\simeq z^4,\qquad \Delta_{i\pm}=2\pm\sqrt{4+m_i^2},\label{eq:UV}
\end{gather}
where $\mu$ is the quark chemical potential, $\muB=N_c\mu$. Since $U(1)_B$ is to be broken spontaneously, we set $\psi_i^{(s)}=0$, so that $\Op_i$ has dimension $\Delta_i\equiv\Delta_{i+}$ ~\cite{Klebanov:1999tb}

\subsection{Charges and dimensions}\label{sec:charges}

The charge under $A_\mu$ is quark number, so a scalar of charge $q$ is dual to an operator containing a net number $q$ of quark fields and baryon number $q/N_c$. Two values are physically distinguished. For $q=2$ the operator is a quark bilinear, the order parameter of color superconductivity~\cite{Alford:2007xm}. A local diquark is a color antitriplet rather than a singlet; as in~\cite{Singh:2025ogp}, its color structure is represented only through an overall factor of $N_c$ in the scalar action, which expresses that all colors condense equally. For $q=2N_c=6$ the operator carries baryon number $2$. It is bosonic and can be a color singlet, and it describes a pair of baryons, the order parameter of neutron superfluidity or proton superconductivity~\cite{Sedrakian:2018ydt}. The same quantum numbers belong to the gauge-invariant order parameter of the CFL phase~\cite{Alford:2007xm,Schafer:1998ef}, so a $q=6$ operator can interpolate between baryon pairing at lower density and color-flavor locking at higher density.

Intermediate charges are not realized by condensing operators in QCD. For $q=3$ the color-singlet $\epsilon_{abc}qqq$ is fermionic and cannot acquire a homogeneous expectation value, which is why baryons must pair before they condense \cite{Sedrakian:2018ydt}. For $q=4$, no color singlet can be formed. We use $q_2=3$ and $4$ only as an interpolation between the physical endpoints $q_2=2$ and $6$, which lets us follow the competition continuously in the charge.

The dimensions are less constrained. Perturbatively an operator with $q$ quark fields has $\Delta=3q/2$, i.e. $\Delta=3, 4.5$, $6$ and $9$ for $q=2$, $3$, $4$ and $6$, respectively. In a strongly interacting medium such operators are strongly renormalized~\cite{Singh:2025ogp}. For $q=6$, a dimension near $3$ would describe a tightly bound bosonic pair with intact constituents, while $\Delta\to3N_c$ would describe a loosely bound, nearly dissolved state. We therefore treat $\Delta_2$, equivalently $m_2^2$, as a free parameter and ask for which values $\Op_2$ condenses together with $\Op_1$.

\section{Boundary conditions and single-condensate phases}\label{sec:background}

In this section we collect the solutions without condensates, the boundary
conditions and physical scales used throughout the paper, and the prescription
that fixes the infrared values of the scalars. We illustrate the prescription
with the single-condensate results of ~\cite{Singh:2025ogp}.

\subsection{Solutions without condensates}\label{sec:nocond}

For $\psi_i=0$, Eqs.~\eqref{eq:EOMg}--\eqref{eq:EOMphi} can be integrated in
closed form, and there are three solutions~\cite{Singh:2024amm}:
\begin{enumerate}
\item Thermal AdS, $g=z^2$, $h=z^4$, $\phi=\mu$, which carries no charge.
\item The charged black hole (CBH),
\begin{gather}
g=z^2\left(1-cz^4+\frac{\zeta\mu^2}{z_H^4}z^6\right), \qquad
h=z^4,\qquad \phi=\mu\left(1-\frac{z^2}{z_H^2}\right),\label{eq:CBH}
\end{gather}
with $c$ fixed by $g(z_H)=0$. It is dual to a deconfined plasma.
\item The charged AdS (CAdS) geometry,
\begin{equation}\label{eq:CAdS}
g=z^2\left(1-cz^4+\zeta Q^2z^6\right),\qquad h=z^4,\qquad \phi=\mu-Qz^2,
\end{equation}
which has no horizon for $\zeta^2Q^4/c^3>4/27$ and is dual to a confined phase at
finite density.
\end{enumerate}
The charge of the CAdS geometry is carried by sources behind the infrared cutoff,
such as wrapped branes or strings, which are known to be pulled into the deep
infrared of AdS~\cite{Brandhuber:1998xy,Gorsky:2015pra}. Its
parameters $Q$ and $c$ are fixed by the conditions at the wall, discussed next.

\subsection{Boundary conditions and scales}\label{sec:bc}

The system \eqref{eq:EOMg}--\eqref{eq:EOMpsi} has two second-order scalar
equations, one second-order equation for $\phi$ and two first-order metric
equations, so a solution requires eight conditions. Four are imposed at the
ultraviolet cutoff $z=\epsilon$: $h(\epsilon)=1$, which fixes the normalization
of boundary time, $\phi(\epsilon)=\mu$, and $\psi_i^{(s)}=0$ for both scalars, so
that $U(1)_B$ is broken spontaneously. We use $\epsilon=10^{-4}$, except when the
second operator has charge $q_2=6$, where we use $\epsilon=10^{-2}$;
Appendix~\ref{app:cutoff} shows that this choice does not affect our conclusions.
The other four conditions are imposed in the infrared and distinguish the phases.

\emph{Deconfined phase.}---At a horizon $z_H$ we require $g(z_H)=0$,
$\phi(z_H)=0$, and regularity of both scalars,
\begin{equation}\label{eq:reg}
\psi_i'(z_H)=\frac{m_i^2}{g'(z_H)}\,\psi_i(z_H),
\end{equation}
where, from Eq.~\eqref{eq:EOMg},
\begin{equation}\label{eq:greg}
g'(z_H)=-4z_H+\frac{2\zeta z_H}{L^2}\left[\frac{h\phi'^2}{4}+\sum_i\frac{g_5^2m_i^2L\psi_i^2}{2}\right]_{z_H}.
\end{equation}
The two scalars are therefore coupled at the horizon through $g'(z_H)$. The
temperature is $T=|g'(z_H)|/(4\pi\sqrt{h(z_H)})$.

\emph{Confined phase.}---At the hardwall $z=z_0$ we set $g(z_0)=g_0=6 z_0^2$, a value
fixed by the $\rho$-meson spectrum~\cite{Singh:2024amm}, and impose the simplest
condition on the gauge field, $\phi(z_0)=0$. In the CAdS solution this gives
$Q=\mu/z_0^2$, and $c$ follows from $g(z_0)=g_0$. The infrared values of the
scalars, $\psi_{i0}\equiv\psi_i(z_0)$, are not fixed by any condition at the
wall; they are determined dynamically, as described in Sec.~\ref{sec:ir}.

In Refs.~\cite{Singh:2024amm,Singh:2025ogp} we also used phenomenological
infrared conditions, in which $\phi'(z_0)$ is fixed by Gauss's law in terms of a
quark density supplied by a van der Waals or NJL equation of state. Here we keep
$\phi(z_0)=0$, the simplest choice for a structural question. With this
condition, the density obtained from $\partial p/\partial\muB$ differs from the
charge read off from the near-boundary expansion of the gauge field,
Eq.~\eqref{eq:UV}. The phenomenological conditions, which tie the data at the
wall to an equation of state for the sources behind it, remove this mismatch. We
therefore use the pressure only to compare solutions at the same $\muB$ and $T$.

\emph{Physical units.}---In the numerical analysis we set $z_0=1$, so that all quantities are measured in units of $z_0^{-1}$. The position of the hardwall is fixed by the glueball mass computed in the absence of flavor degrees of freedom~\cite{Singh:2024amm}, which gives $z_0^{-1}=290$~MeV. A quantity of mass dimension $n$ obtained from the numerical solutions is converted to physical units by multiplying it by $z_0^{-n}$. In particular, the quark chemical potential $\mu$ and the temperature are multiplied by $z_0^{-1}$, and the baryon chemical potential is $\muB=N_c\mu z_0^{-1}$ with $N_c=3$. For example, $\mu=2.2$, the value used in Sec.~\ref{sec:cads}, corresponds to $\muB\simeq1.91$~GeV.

\subsection{Infrared amplitudes and a single condensate}\label{sec:ir}

Once the scalars are switched on, nothing at the wall fixes $\psi_{i0}$. In
\cite{Singh:2025ogp} we fixed it by extremizing the on-shell action, which in
the grand canonical ensemble amounts to maximizing the pressure. The reasoning is
as follows. The on-shell action can be written as a total
derivative \cite{Sonner:2010yx}, and the contribution depends only on $g$, $h$, $\phi$ and $\phi'$ at the
wall and not on the scalar there. This remains true if the region $z>z_0$ contains
a condensate of its own. If $z_0$ is regarded as an ultraviolet cutoff for that
region, $\psi_{i0}$ is a free parameter of the effective description, and the
natural choice is the one that extremizes the action. The prescription makes the
onset of condensation continuous, since the selected amplitude grows from zero.
At the densities considered here, it also selects solutions away from the curve on
which $g_{tt}(z_0)\to0$, a naked singularity that ends the family of solutions as
the amplitudes grow. For small
amplitudes the scalar equation defines a Sturm--Liouville problem whose lowest
mode has no node, and a node that appears when the solution is continued beyond
the wall is expected to survive in any infrared
completion~\cite{Singh:2025ogp}. Therefore, we require the profiles to be nodeless.

\begin{figure}[h]
\centering
\subfloat[$\psi(z)$\label{fig:single_sol}]{\figorplaceholder{0.45\linewidth}{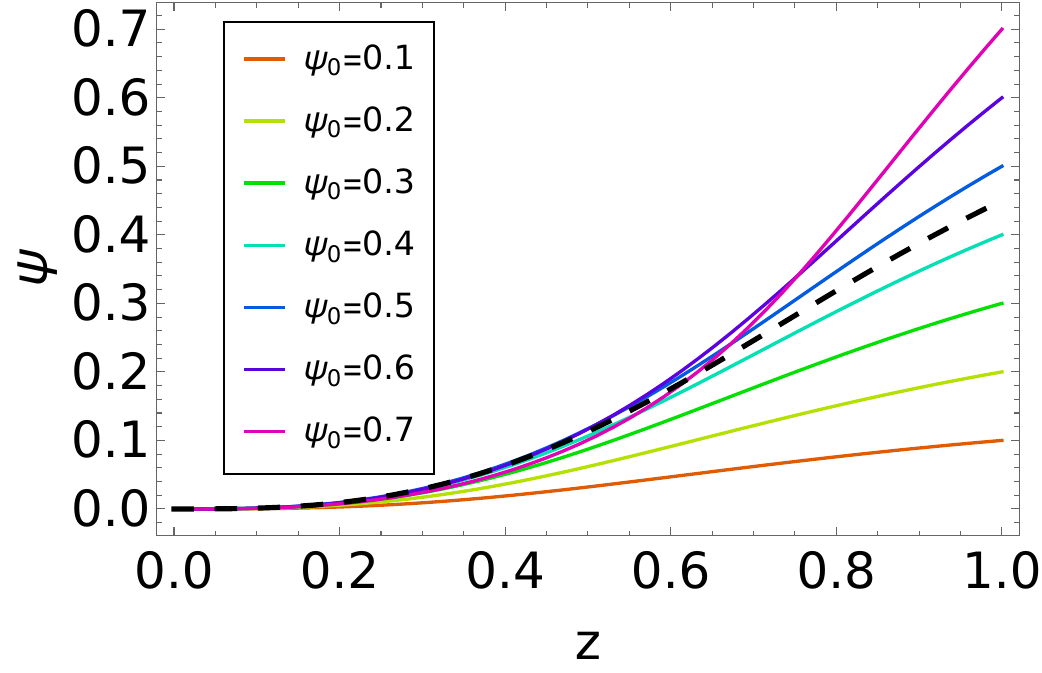}}\hfill
\subfloat[Pressure vs $\psi_0$\label{fig:single_p}]{\figorplaceholder{0.45\linewidth}{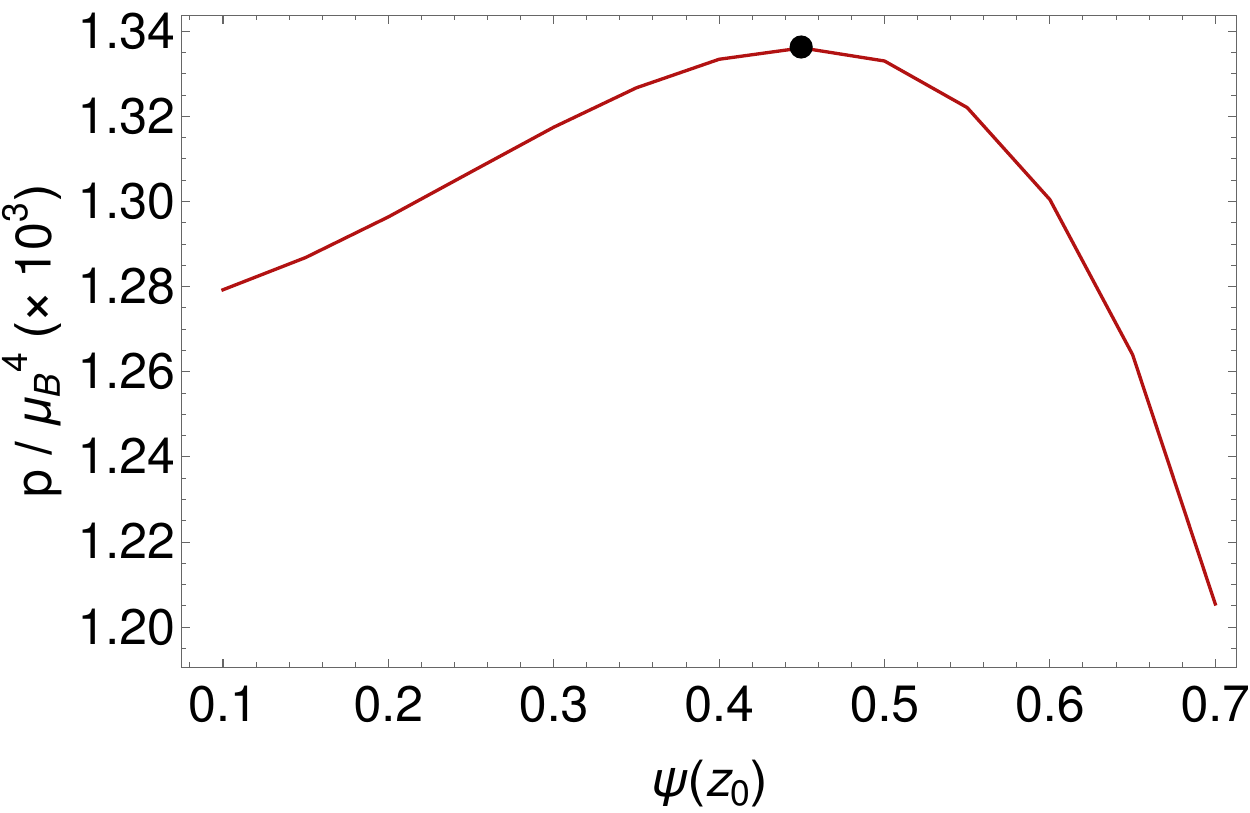}}\\
\subfloat[Selected $\psi_0$\label{fig:single_y0}]{\figorplaceholder{0.45\linewidth}{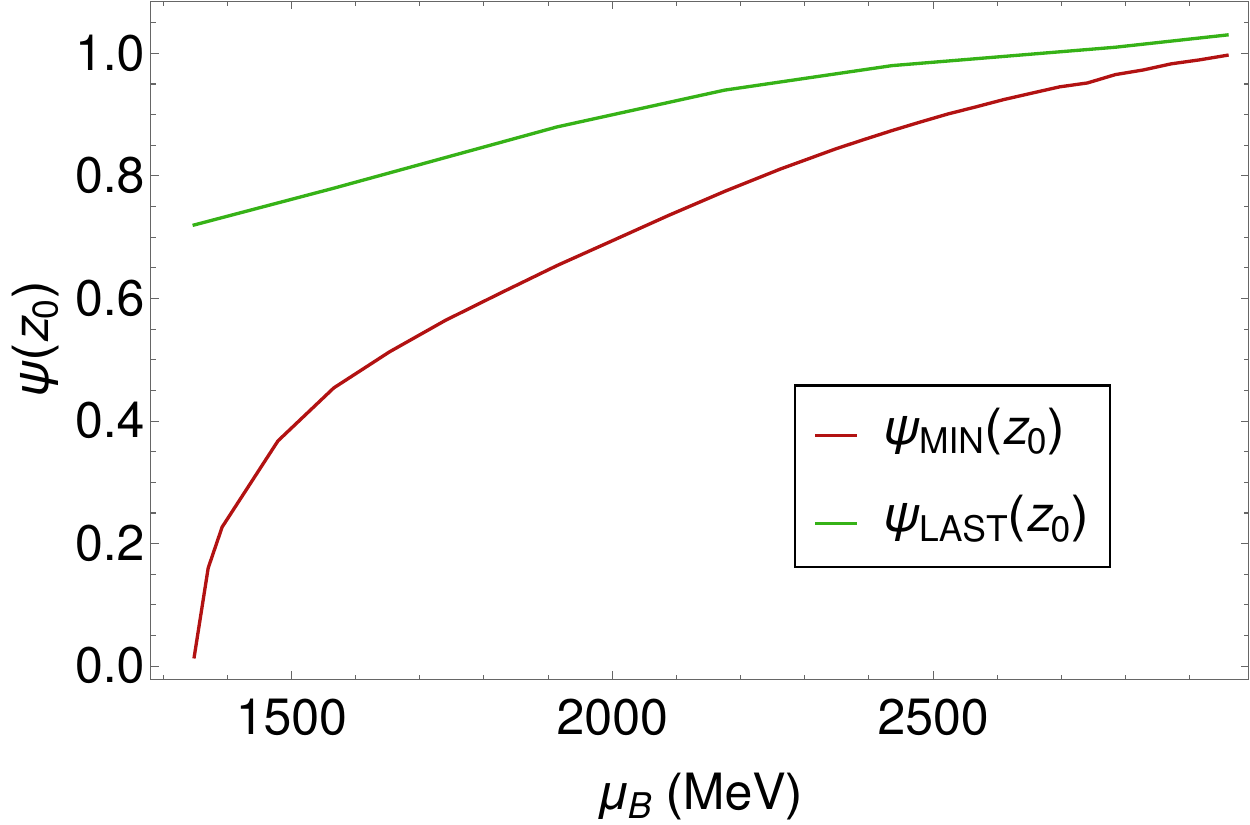}}\hfill
\subfloat[Phase diagram\label{fig:single_pd}]{\IfFileExists{Figures/PD_Phi0_q2_zeta0p7.pdf}{%
  \resizebox{0.45\linewidth}{!}{\begin{overpic}[scale=0.25,unit=1mm]{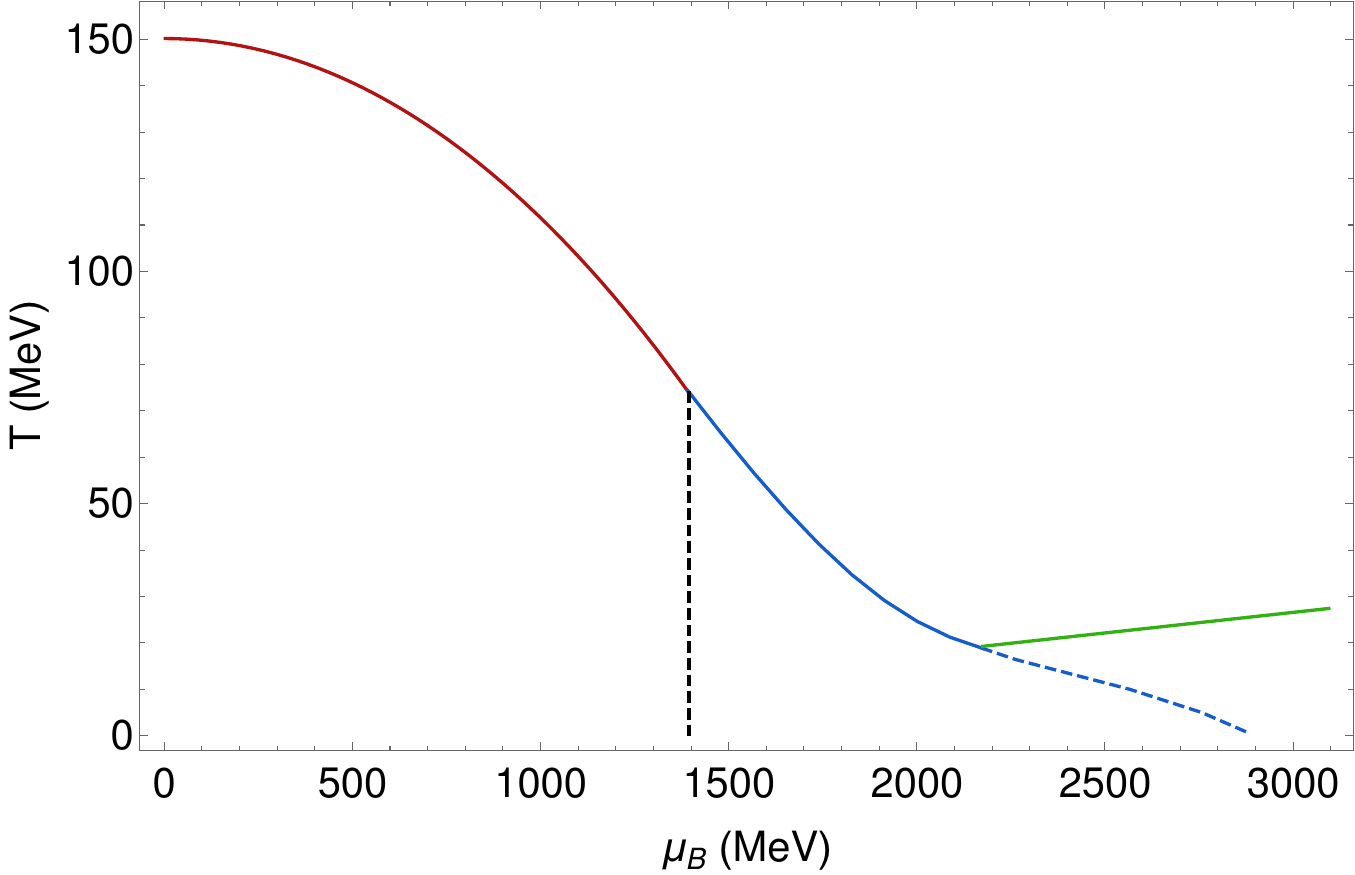}
  \tiny\put(28.5,33){CAdS}\put(55,12){CAdS+$\Delta$}\put(60,40){CBH}\put(82,14.5){CBH+$\Delta$}
  \end{overpic}}}{\figplaceholder{0.45\linewidth}{2.8cm}{add \texttt{Figures/PD\_Phi0\_q2\_zeta0p7.pdf}\\(from ~\cite{Singh:2025ogp})}}}
\caption{Single-condensate results of ~\cite{Singh:2025ogp} for $q=2$,
$\Delta=3$, $\zeta=0.77$ and $\phi(z_0)=0$, reproduced under the CC~BY~4.0
license. (a)~Scalar profiles for several $\psi_0$ at $\muB=1.56$~GeV; the dashed
curve is the maximum-pressure solution. (b)~Reduced pressure as a function of
$\psi_0$ at the same $\muB$; the dot marks the maximum. (c)~Selected $\psi_0$ as a
function of $\muB$ (red) and the value at which $g_{tt}(z_0)\to0$ (green).
(d)~Schematic phase diagram; ``$+\Delta$'' denotes the condensed phases.}
\label{fig:single}
\end{figure}

Figure~\ref{fig:single} summarizes how this works for a single scalar with $q=2$, $\Delta=3$ and $\zeta=0.77$~\cite{Singh:2025ogp}. At fixed $\muB$ there is a one-parameter family of sourceless, nodeless solutions labelled by $\psi_0$, Fig.~\ref{fig:single_sol}. As $\psi_0$ grows, $h(z_0)$ increases and $g_{tt}(z_0)$ decreases towards zero, which bounds the family from above (green curve in Fig.~\ref{fig:single_y0}; the same behavior for two scalars is shown in Fig.~\ref{fig:sol_gtt}). The pressure has a maximum at an intermediate value of $\psi_0$, well inside this bound, Fig.~\ref{fig:single_p}. At the onset the selected $\psi_0$ grows continuously from zero, and at large density it approaches the singular branch,  Fig.~\ref{fig:single_y0}. However, before reaching the singular branch, the deconfined phase with condensate becomes the thermodynamically preferred phase. Thus, we never see a solution with a naked singularity.

In the $\muB$--$T$ plane, Fig.~\ref{fig:single_pd} shows that the onset in the confined
phase is a vertical line, because with $\phi(z_0)=0$ the pressure of the
horizonless solutions does not depend on temperature. At larger $\muB$ the
confined condensed phase makes a first-order transition to the charged black hole
with condensate, and at high temperature the uncondensed black hole is preferred.

\section{Two condensates in the confined phase}\label{sec:cads}

Having discussed the single-condensate case, we now turn to configurations in which two charged scalar fields are simultaneously nonzero. Throughout, we fix $\Op_1$ to be the diquark-like operator, with charge $q_1 = 2$ and scaling dimension $\Delta_1 = 3$, and vary the charge $q_2$ and scaling dimension $\Delta_2$ of the second operator $\Op_2$. This allows us to determine which operator condenses first and whether the two condensates can coexist.

\subsection{Solutions}

\begin{figure}[h]
\centering
\subfloat[$\psi_1(z)$\label{fig:sol_y1}]{\includegraphics[width=0.45\linewidth]{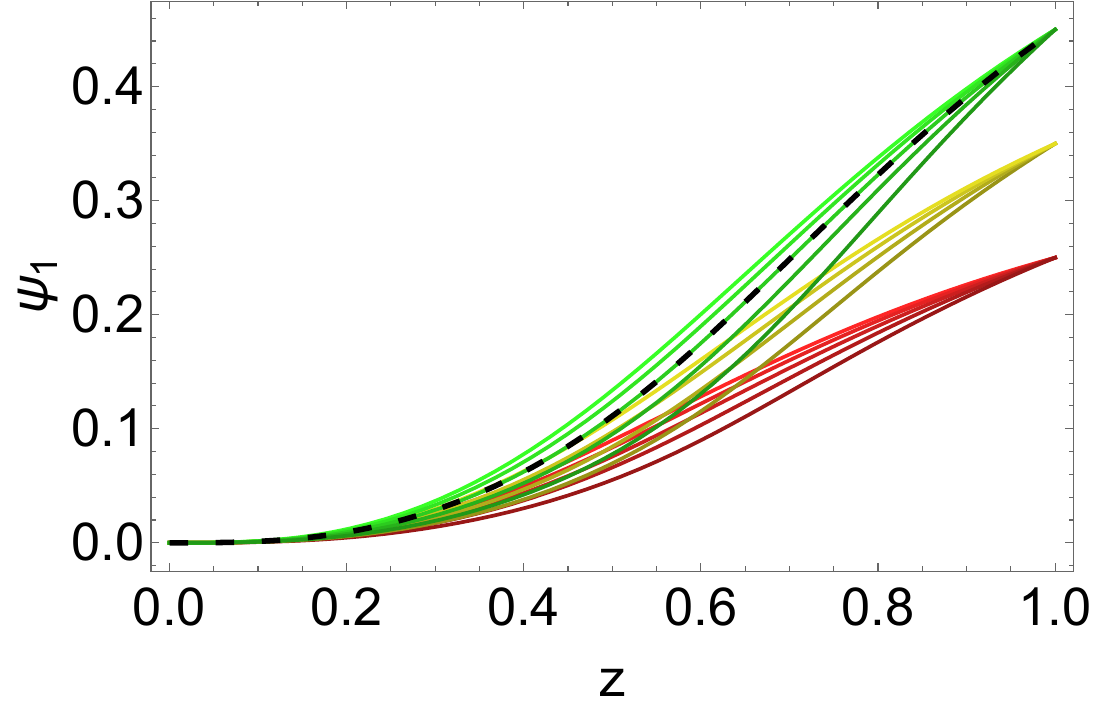}}\hfill
\subfloat[$\psi_2(z)$\label{fig:sol_y2}]{\includegraphics[width=0.45\linewidth]{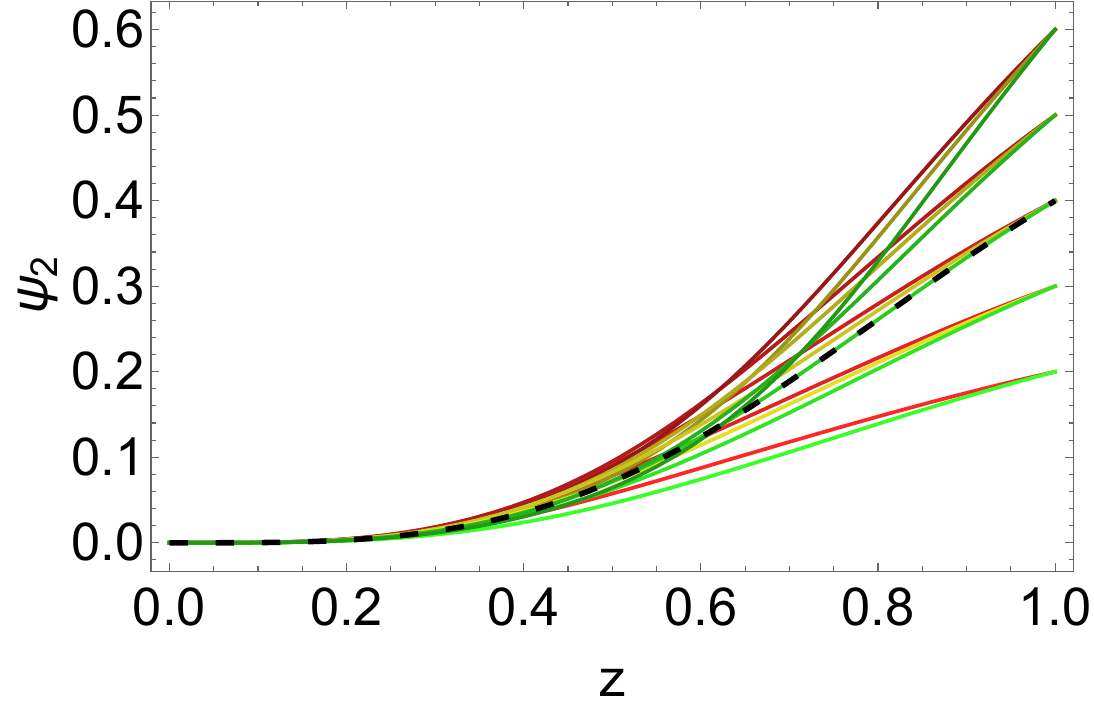}}\\
\subfloat[$\phi'(z)$\label{fig:sol_fp}]{\includegraphics[width=0.45\linewidth]{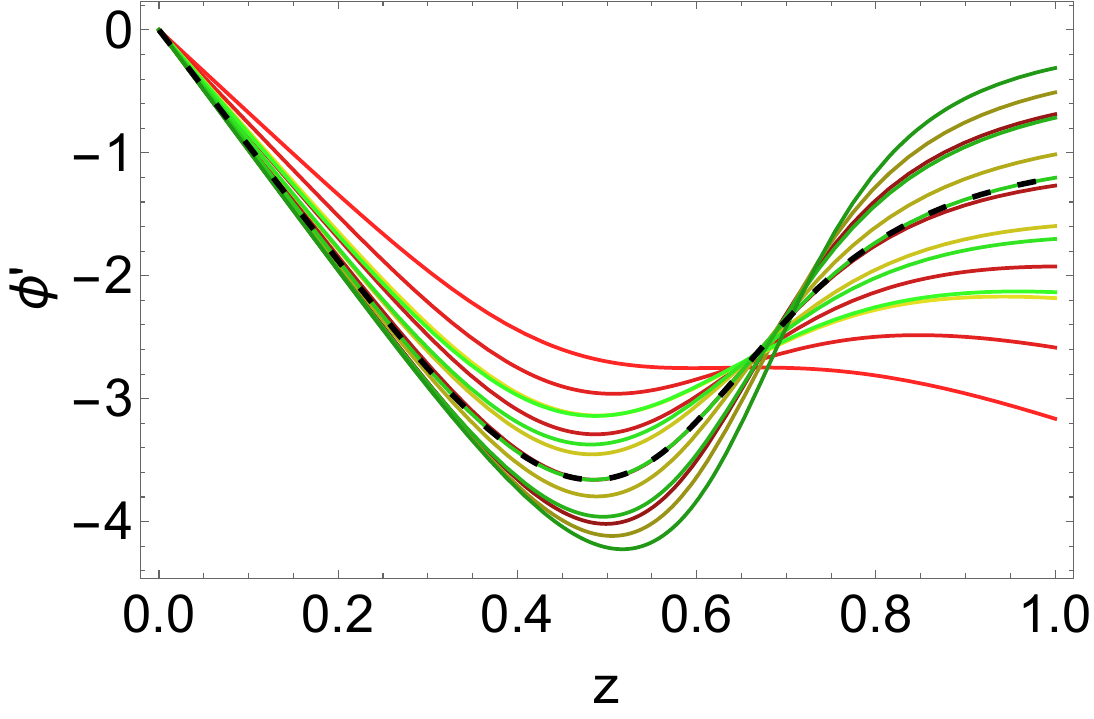}}\hfill
\subfloat[$g_{tt}(z)$\label{fig:sol_gtt}]{\includegraphics[width=0.45\linewidth]{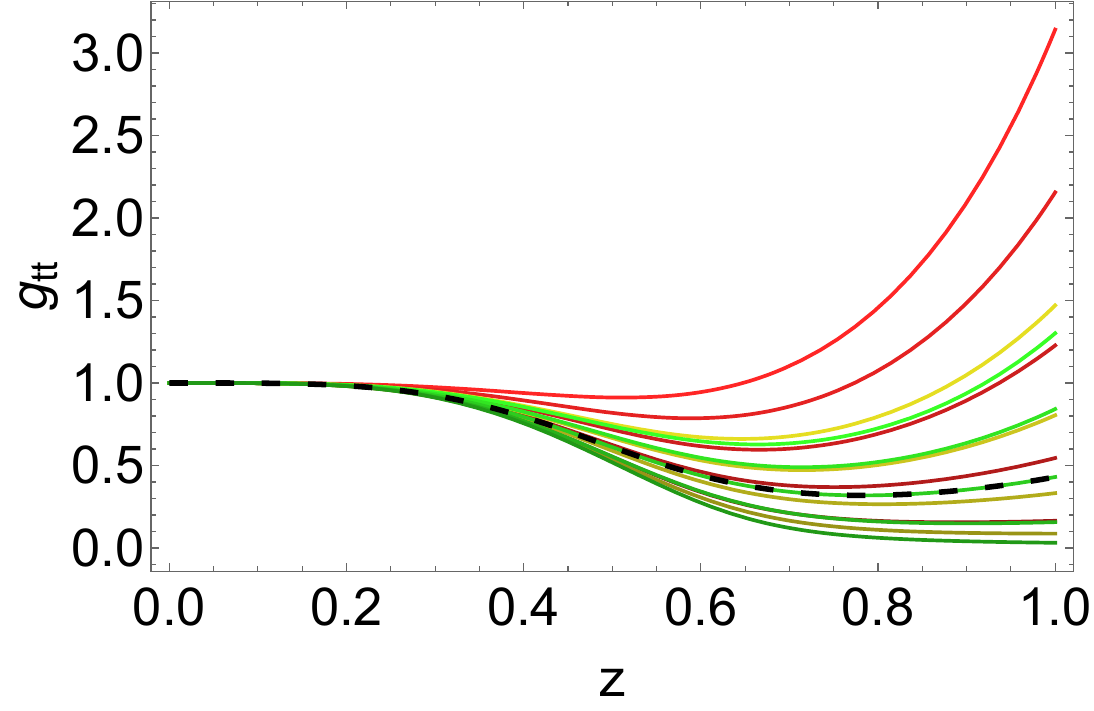}}
\caption{Two-field CAdS solutions at $\mu=2.2$ for $q_1=2$, $\Delta_1=3$ and $q_2=3$, $m_2^2=-0.5$ ($\Delta_2\simeq3.87$), with $\zeta=0.77$. The hue labels $\psi_{10}$ (red $0.25$, yellow $0.35$, green $0.45$) and the brightness labels $\psi_{20}$ (from $0.2$ to $0.6$, darkest is the largest). Black dashed curves: maximum-pressure solution, $(\psi_{10},\psi_{20})\simeq(0.48,0.38)$.}
\label{fig:sol}
\end{figure}

We first show that well-behaved two-field solutions exist for a range of IR boundary values/ infrared amplitudes ($\psi_{10},\psi_{20}$), among which the maximum-pressure condition selects one. Figure~\ref{fig:sol} shows CAdS solutions at $\mu=2.2$ for $q_2=3$, $m_2^2=-0.5$ ($\Delta_2\simeq3.87$) on a grid of infrared amplitudes. The hue labels $\psi_{10}$ (red $0.25$, yellow $0.35$, green $0.45$) and the brightness labels $\psi_{20}$ (from $0.2$ to $0.6$, darkest=largest). The black dashed curves represent the maximum-pressure solutions at $(\psi_{10},\psi_{20})\simeq(0.48,0.38)$. 

All profiles are smooth and nodeless. At fixed $\psi_{10}$ the profile of $\psi_1$ is lowered slightly as $\psi_{20}$ grows, and vice versa [Figs.~\ref{fig:sol_y1} and \ref{fig:sol_y2}]. The electric field shows how the charge is redistributed [Fig.~\ref{fig:sol_fp}]. As either amplitude grows, $|\phi'|$ increases in the middle of the bulk and decreases near the wall. Since the flux through the wall measures the charge of the sources behind it, the condensates take over part of the charge that would otherwise be carried by that region. The metric component $g_{tt}$ decreases near the wall as the amplitudes grow [Fig.~\ref{fig:sol_gtt}], as in the single-field case. The family of solutions ends where $g_{tt}(z_0)\to0$, which is now a curve in the $(\psi_{10},\psi_{20})$ plane rather than a single value of $\psi_0$. 

We did not use the perturbative value $\Delta_2=4.5$ here because it is numerically stiff at small $\epsilon$, and, as shown below, the pressure maximum does not contain both condensates at that dimension.

\subsection{Selection by maximum pressure}

\begin{figure}[h]
\centering
\subfloat[$m_2^2=0$ ($\Delta_2=4$): only $\Op_1$\label{fig:cont_a}]{\includegraphics[width=0.32\linewidth]{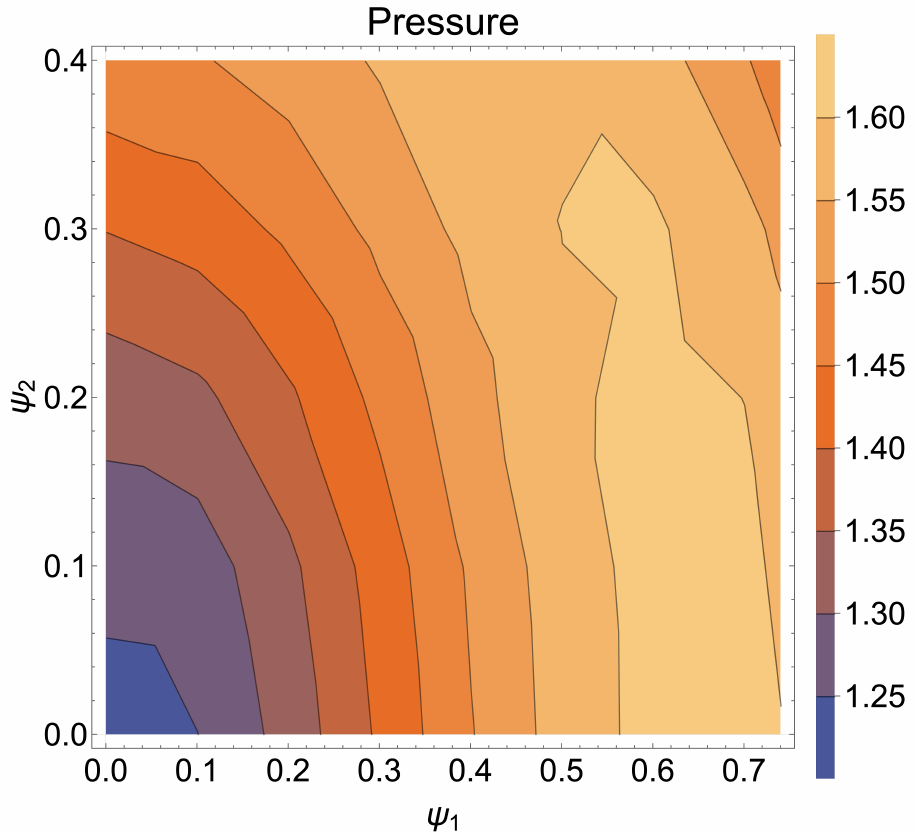}}\hfill
\subfloat[$m_2^2=-0.5$ ($\Delta_2\simeq3.87$): both\label{fig:cont_b}]{\includegraphics[width=0.32\linewidth]{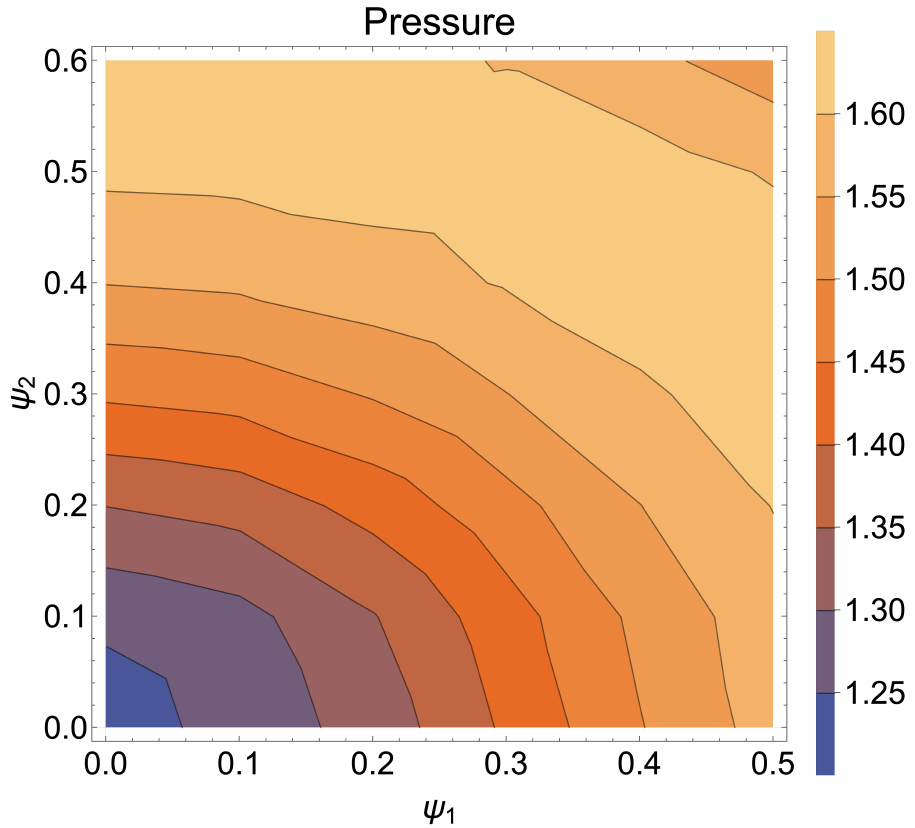}}\hfill
\subfloat[$m_2^2=-1$ ($\Delta_2\simeq3.73$): only $\Op_2$\label{fig:cont_c}]{\includegraphics[width=0.32\linewidth]{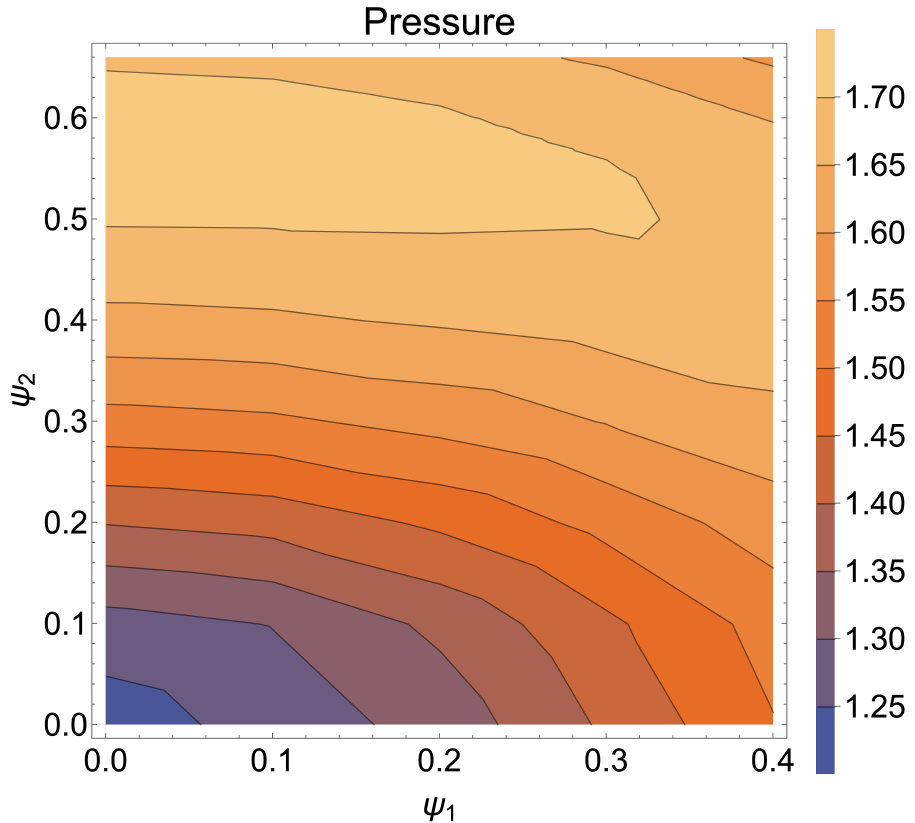}}
\caption{Pressure over the plane of infrared amplitudes at $\mu=2.2$ for $q_1=2$, $\Delta_1=3$, $q_2=3$ and $\zeta=0.77$, from blue (lowest) to beige (highest).} 
\label{fig:contours}
\end{figure}

Figure~\ref{fig:contours} shows the pressure over the $(\psi_{10},\psi_{20})$ plane for three values of $m_2^2$ at $q_2=3$. At $m_2^2=0$ ($\Delta_2=4$) the maximum lies on the axis $\psi_{20}=0$ and only $\Op_1$ condenses. At $m_2^2=-1$ ($\Delta_2\simeq3.73$) it lies on $\psi_{10}=0$ and only $\Op_2$ condenses. At $m_2^2=-0.5$ the maximum is an interior point, $(\psi_{10},\psi_{20})\simeq(0.48,0.38)$, and the two operators condense together. The contours around the maximum are elongated. The pressure changes very little along the direction in which $\psi_{10}$ decreases and $\psi_{20}$ increases, i.e.\ when one condensate is traded for the other. As a consequence, the location of the maximum is more sensitive to numerical errors than its height (Appendix~\ref{app:cutoff}).

\begin{figure}[h]
\centering
\includegraphics[width=0.5\linewidth]{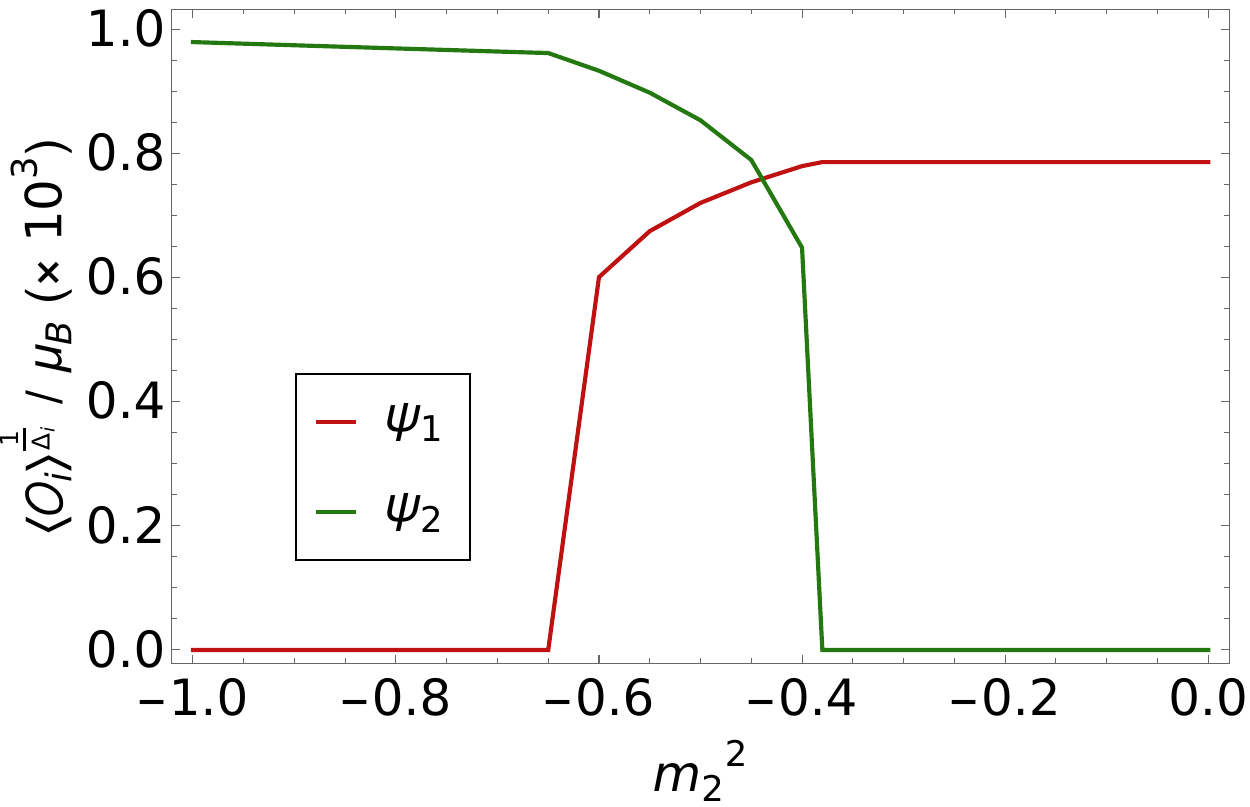}
\caption{Condensates $\vev{\Op_i}^{1/\Delta_i}/\muB$ at the pressure maximum as functions of $m_2^2$, for $q_2=3$ at $\mu=2.2$. Both are nonzero for $-0.650<m_2^2<-0.388$.}
\label{fig:Ovsm2}
\end{figure}

The pattern follows from Eq.~\eqref{eq:meff}. At fixed charge, lowering $m_2^2$ makes $\Op_2$ easier to condense, so for small enough $\Delta_2$ it wins; for large enough $\Delta_2$ it is $\Op_1$ that wins. Between the two there is a range in which neither wins outright. Figure~\ref{fig:Ovsm2} shows the condensates as functions of $m_2^2$ at fixed $\mu=2.2$. Coexistence occurs for $-0.650<m_2^2<-0.388$. Inside this range the $\Op_1$ condensate decreases and the $\Op_2$ condensate increases as $m_2^2$ is lowered, and each vanishes at one edge of the window.  
Because we plot $\langle\Op_i\rangle^{1/\Delta_i}$, the onset and disappearance of the condensates appear steeper than they are.

To characterize the transitions more precisely, Table~\ref{tab:amps} lists the infrared amplitudes $\psi_{10}$ and $\psi_{20}$ of the solution that maximizes the pressure at each point in the window. The squared amplitudes vanish linearly at both edges, as expected for continuous transitions. A linear fit to the interior points gives $\psi_{10}^2=0$ at $m_2^2=-0.643$ and $\psi_{20}^2=0$ at $m_2^2=-0.381$, within $0.01$ of the edges found from the maximization.

\begin{table}[h]
\caption{Infrared amplitudes at the pressure maximum for $q_2=3$ at $\mu=2.2$. The coexistence window is $-0.650<m_2^2<-0.388$.}
\label{tab:amps}
\begin{ruledtabular}
\begin{tabular}{cccc}
$m_2^2$ & $\Delta_2$ & phase & $(\psi_{10},\psi_{20})$\\
\hline
$-1.00$ & $3.73$ & only $\Op_2$ & $(0,\,0.57)$\\
$-0.60$ & $3.84$ & both & $(0.27,\,0.51)$\\
$-0.55$ & $3.86$ & both & $(0.39,\,0.46)$\\
$-0.50$ & $3.87$ & both & $(0.48,\,0.38)$\\
$-0.45$ & $3.88$ & both & $(0.56,\,0.29)$\\
$-0.42$ & $3.89$ & both & $(0.61,\,0.21)$\\
$\phantom{-}0.00$ & $4.00$ & only $\Op_1$ & $(0.65,\,0)$\\
\end{tabular}
\end{ruledtabular}
\end{table}

\subsection{Coexistence windows in the operator plane}

Repeating the scan for $q_2=4$ and $6$ gives the windows
\begin{align}
q_2=3:&\; -0.650<m_2^2<-0.388, & 3.830<\Delta_2<3.901,\nonumber\\
q_2=4:&\; \phantom{-}2.257<m_2^2<3.369, & 4.501<\Delta_2<4.715,\label{eq:windows}\\
q_2=6:&\; \phantom{-}9.633<m_2^2<13.579, & 5.692<\Delta_2<6.193,\nonumber
\end{align}
all at $\mu=2.2$. The edges are the values of $m_2^2$ at which the pressure maximum reaches one of the axes. Two trends are clear. First, the window moves to larger $\Delta_2$ as $q_2$ increases. An operator with a larger charge can afford to be heavier and still compete, since the charge term in Eq.~\eqref{eq:meff} grows as $q_2^2$. Second, the window also widens, from $0.07$ in $\Delta_2$ at $q_2=3$ to $0.21$ at $q_2=4$ and $0.50$ at $q_2=6$. In all three cases the perturbative dimension $3q_2/2$ lies well above the window, so a coexisting phase requires an operator whose dimension is strongly reduced from its free-field value.

\begin{figure}[t]
\centering
\includegraphics[width=0.5\linewidth]{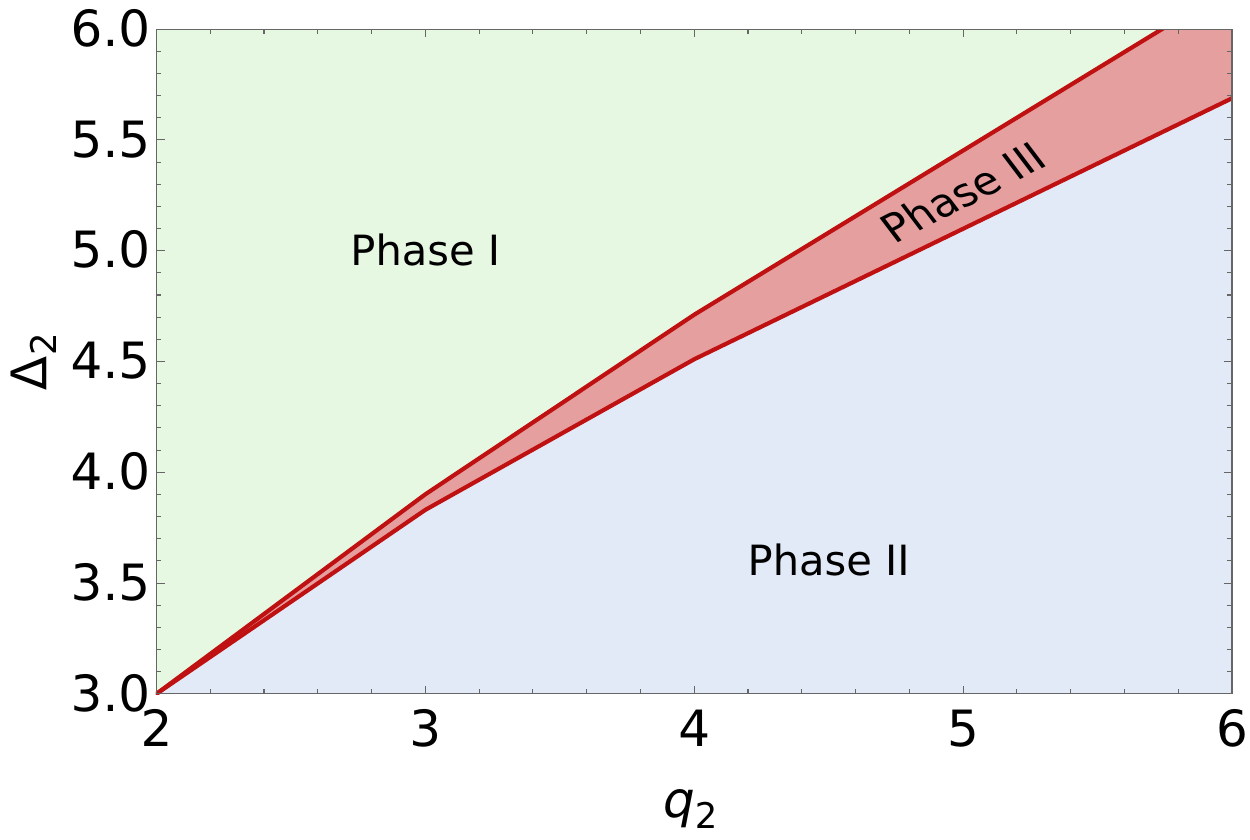}
\caption{Phases in the plane of operator data $(q_2,\Delta_2)$ at $q_1=2$, $\Delta_1=3$, $\mu=2.2$ and $\zeta=0.77$. Phase~I: only $\Op_1$ condenses. Phase~II: only $\Op_2$ condenses. Phase~III: coexistence. }
\label{fig:opplane}
\end{figure}

Figure~\ref{fig:opplane} collects these results in the $(q_2,\Delta_2)$ plane at fixed $q_1$, $\Delta_1$, $\mu$ and $\zeta$. Phase~I, at large $\Delta_2$, is the region in which only $\Op_1$ condenses. Phase~II, at small $\Delta_2$, the region in which only $\Op_2$ condenses. Phase~III, the strip between them, the region of coexistence. At $q_2=2$ the strip closes at $\Delta_2=3$, where $\Op_2$ coincides with $\Op_1$. Two intermediate charges at the same $\mu$ confirm the interpolation. For $(q_2,m_2^2)=(4.5,4.5)$ and $(5,6.25)$, i.e.\ $\Delta_2\simeq4.92$ and $5.20$, the pressure maximum is an interior point of the $(\psi_{10},\psi_{20})$ plane, and both points lie inside the strip of Fig.~\ref{fig:opplane}.

The narrowness of the windows has a simple interpretation in the bulk. Each
condensate is a cloud of charge distributed along the radial direction, with
density proportional to $q_i^2\sqrt{h}\,\phi\,\psi_i^2/(z^3g)$. Through Gauss's
law, Eq.~\eqref{eq:EOMphi}, each cloud screens the chemical potential seen by the
other. This costs pressure in proportion to the overlap of the two clouds, an
electrostatic repulsion in the AdS bulk. In the confined geometry both clouds are
concentrated in the same region between the middle of the bulk and the wall
(Fig.~\ref{fig:sol}), so the repulsion is strong and coexistence survives only
when the two instabilities are nearly equal in strength. When the operators
coincide, so do the clouds, and the window closes.

In short, at $\mu=2.2$ the maximum-pressure condition selects  both  condensates only within the
narrow strip of Fig.~\ref{fig:opplane}, which widens with $q_2$ and closes at the
degenerate point. These windows were obtained at a single value of the chemical
potential. In Sec.~\ref{sec:thermo} we fix the operator data inside the windows
and follow the confined phase as a function of $\muB$, and in Sec.~\ref{sec:cbh}
we show that the same structure appears in the deconfined phase.

\section{Dependence on the chemical potential}\label{sec:thermo}

In this section we fix the operator data and follow the confined phase as a
function of the baryon chemical potential. We consider two cases, the
interpolating operator $q_2=3$, $m_2^2=-0.5$ ($\Delta_2\simeq3.87$), and the
dibaryon-like operator $q_2=6$, $m_2^2=10$ ($\Delta_2\simeq5.74$), both of which
lie inside the coexistence windows of Eq.~\eqref{eq:windows}. For each we show
the two condensates and the pressure, and compare the latter with the pressure of
CAdS without condensates. With $\phi(z_0)=0$ the horizonless solutions do not
depend on temperature, so these results hold throughout the confined phase.

\emph{Order of the transitions.}---The condensation transitions in the confined
phase are continuous. At each onset the selected infrared amplitude grows from
zero, and the pressure of the condensed phase joins the CAdS pressure
tangentially, so both the pressure and its slope are continuous. The same holds
at the second onset, where the second condensate grows from zero in the presence
of the first. The transitions between the confined and deconfined geometries are
of first order and are discussed in Sec.~\ref{sec:pd}.

\subsection{Interpolating charge, $q_2=3$}

\begin{figure}[h]
\centering
\subfloat[Condensates\label{fig:q3_cond}]{\includegraphics[width=0.49\linewidth]{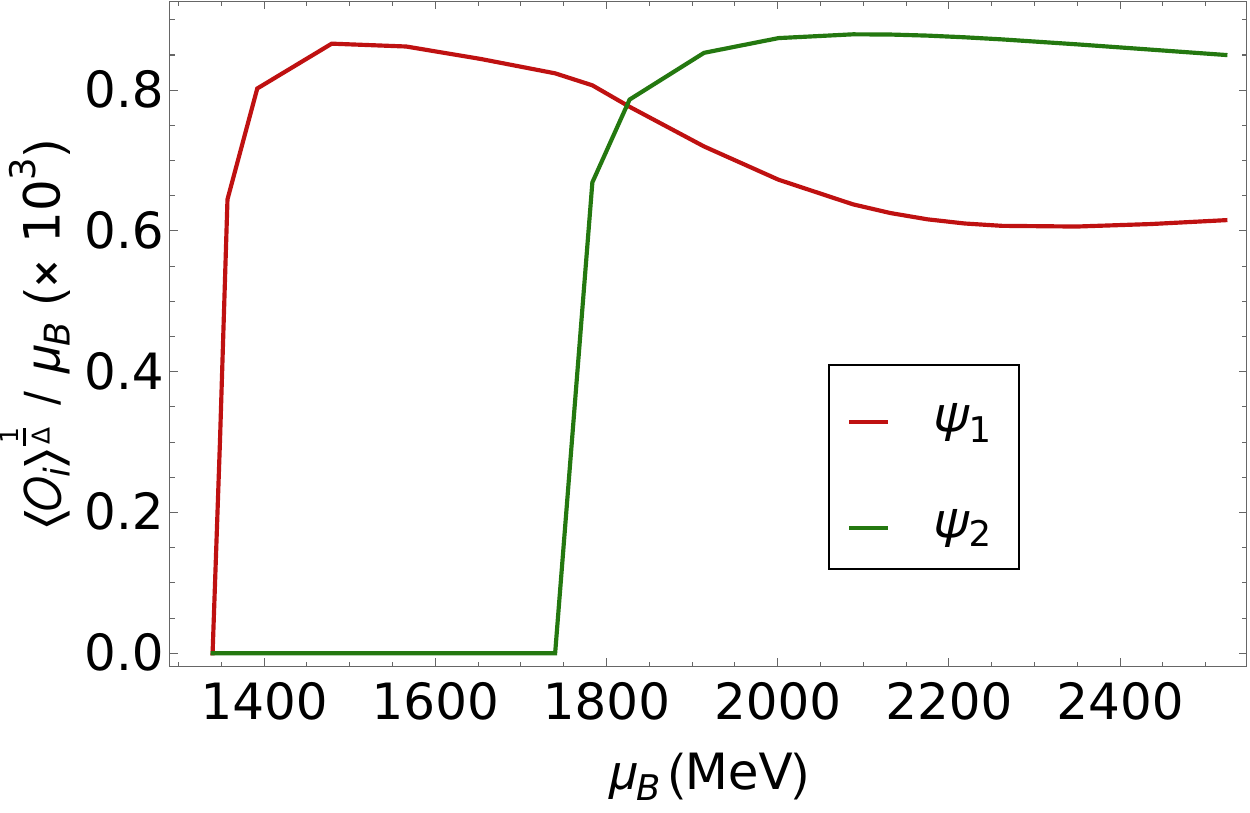}}\hfill
\subfloat[Pressure\label{fig:q3_p}]{\includegraphics[width=0.49\linewidth]{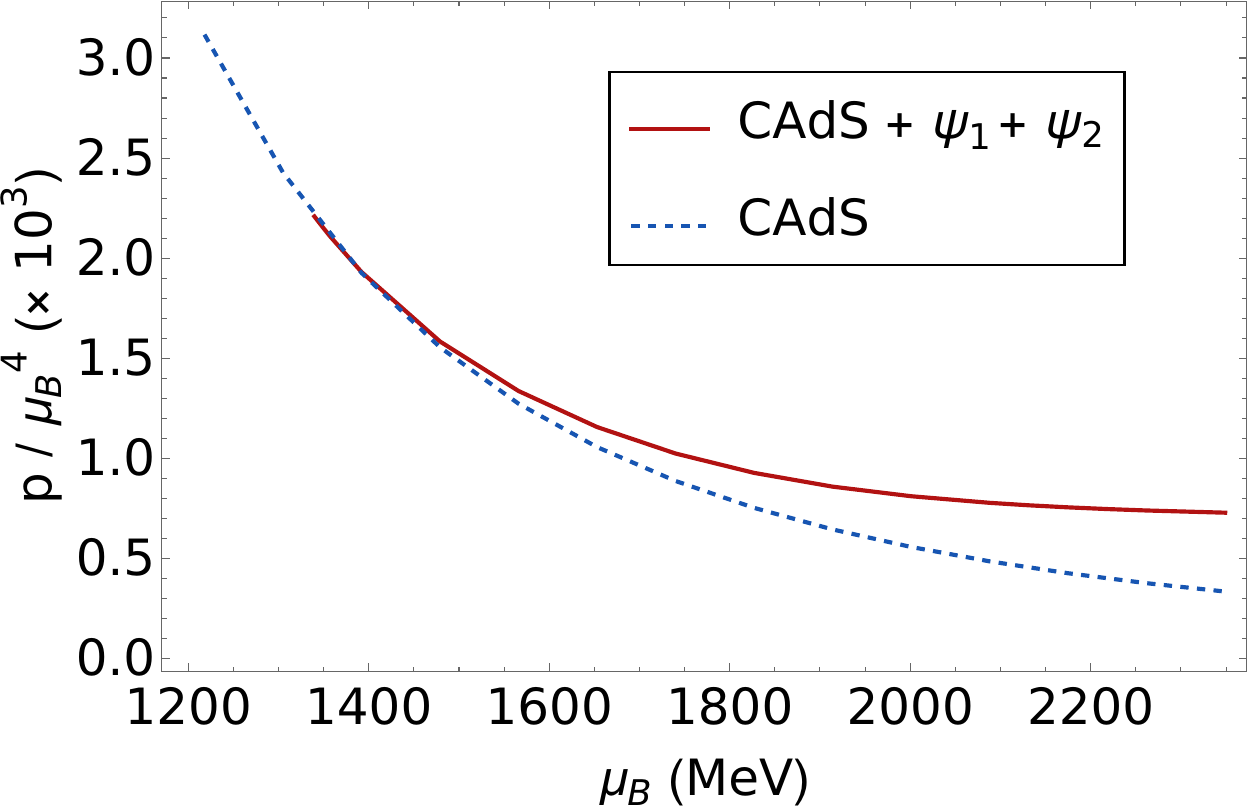}}
\caption{Observables in the confined phase as a function of $\muB$ for $q_1=2$, $m_1^2=-3$ and
$q_2=3$, $m_2^2=-0.5$, at $\zeta=0.77$. (a)~Condensates; the labels $\psi_1$,
$\psi_2$ refer to $\Op_1$, $\Op_2$. (b)~Pressure with (solid) and without
(dashed) condensates.}
\label{fig:thermo_q3}
\end{figure}

For $q_2 = 3$, the diquark-like operator $\Op_1$ condenses first, at $\muB \approx 1.35$~GeV, and $\Op_2$ condenses at $\muB \approx 1.75$~GeV in the presence of $\Op_1$ [Fig.~\ref{fig:q3_cond}; cf.\ Fig.~\ref{fig:pd_q3}]. Just above its onset, the ratio $\vev{\Op_1}^{1/3}/\muB$ rises steeply, reaches a maximum, and then decreases, so at large $\muB$ the condensate grows more slowly than $\muB^3$. It decreases faster once $\Op_2$ appears, indicating that the two condensates compete, and $\vev{\Op_1}^{1/3}/\muB$ eventually levels off. The ratio $\vev{\Op_2}^{1/\Delta_2}/\muB$ passes through a broad maximum and decreases slightly at the largest $\muB$ shown. Neither condensate is driven to zero. Once both are present, they coexist over the entire range we consider. Beyond the first onset, the condensed phase has a higher pressure than the uncondensed phase at every $\muB$, and at $\muB \approx 2.3$~GeV its value of $p/\muB^4$ is about twice that of CAdS without condensates [Fig.~\ref{fig:q3_p}].

\subsection{Diquark and dibaryon, $q_2=6$}

\begin{figure}[h]
\centering
\subfloat[Condensates\label{fig:q6_cond}]{\includegraphics[width=0.49\linewidth]{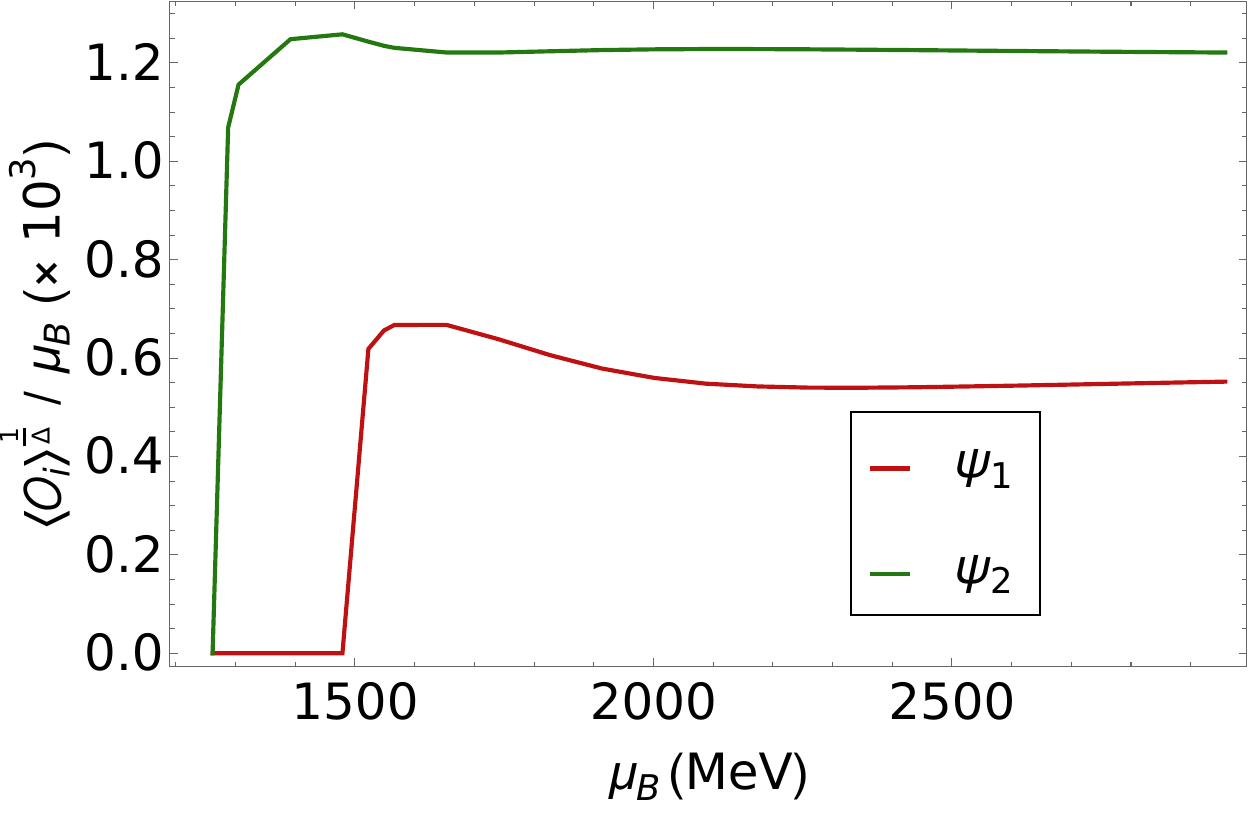}}\hfill
\subfloat[Pressure\label{fig:q6_p}]{\includegraphics[width=0.49\linewidth]{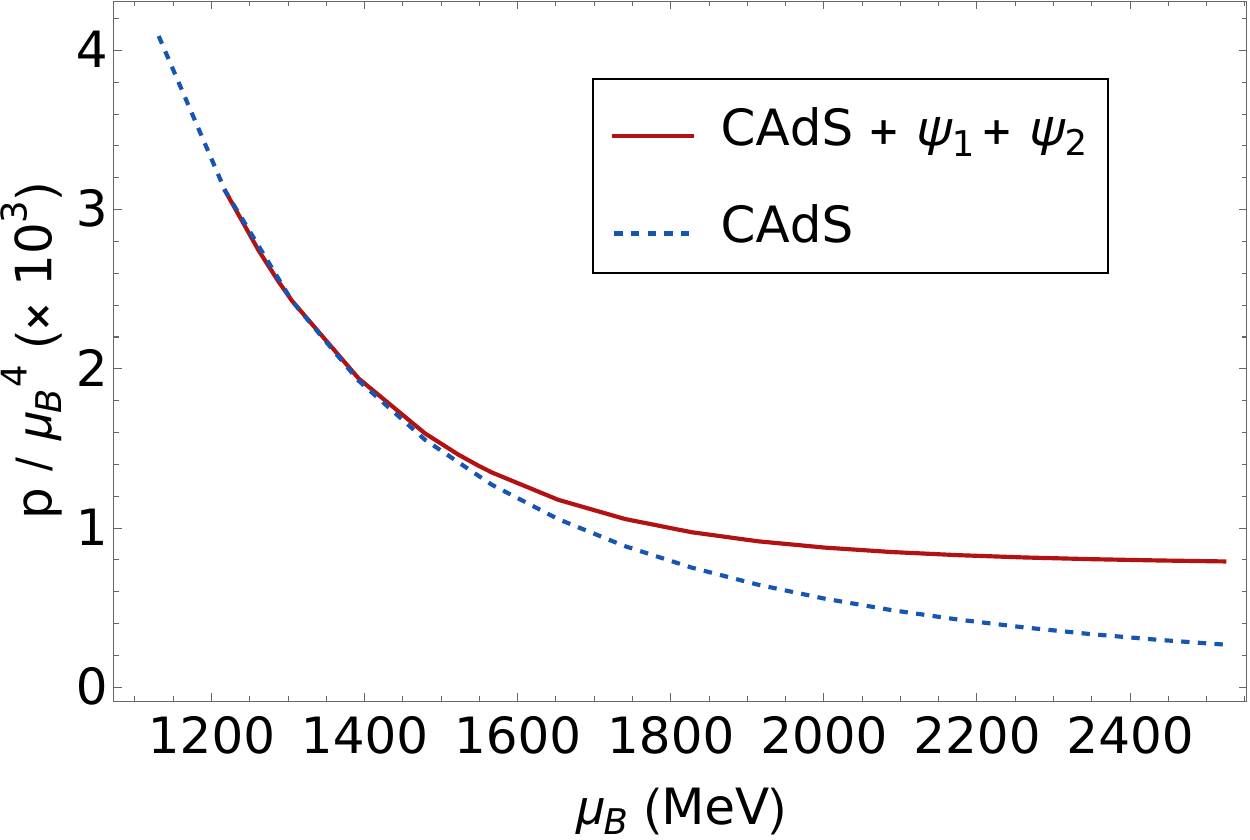}}
\caption{As in Fig.~\ref{fig:thermo_q3}, for the diquark-like operator
($q_1=2$, $m_1^2=-3$) and the dibaryon-like operator ($q_2=6$, $m_2^2=10$,
$\Delta_2\simeq5.74$).}
\label{fig:thermo_q6}
\end{figure}

The physically motivated case is the dibaryon-like operator, $q_2=2N_c=6$. We take
$m_2^2=10$, i.e.\ $\Delta_2=2+\sqrt{14}\simeq5.74$. The order of condensation is
now reversed [Fig.~\ref{fig:q6_cond}]. The dibaryon condenses first, at
$\muB\approx1.26$~GeV, and the diquark joins it at $\muB\approx1.5$~GeV
(cf.\ Fig.~\ref{fig:pd_q6}).
The dibaryon ratio rises steeply and then stays nearly constant, changing only
slightly when the diquark appears; the diquark ratio reaches a maximum shortly
after its onset and then decreases slowly. Once again, when both are present, they persist. Similar condensation sequences, in which both condensates persist once they have formed, were found in black hole backgrounds by Cai \textit{et al.}~\cite{Cai:2013wma}
(their Phases~A and B). Since those backgrounds are deconfined, we make the
comparison in Sec.~\ref{sec:cbh}.


At $\mu=2.2$ ($\muB\simeq1.91$~GeV) the value $m_2^2=10$ lies close to the lower
edge of the window, $m_2^2=9.633$, where the diquark condensate vanishes.
Appendix~\ref{app:cutoff} shows that coexistence at $m_2^2=10$ is stable under
changes of the ultraviolet cutoff for $1.8\le\mu\le2.6$
($1.57\lesssim\muB\lesssim2.26$~GeV).

The condensates raise the pressure considerably; at $\muB\approx2.5$~GeV it is
about three times that of CAdS without condensates [Fig.~\ref{fig:q6_p}]. In the
condensed phase $p/\muB^4$ levels off at large $\muB$, whereas it keeps decreasing
in CAdS without condensates.

\section{Deconfined phase}\label{sec:cbh}

In the deconfined phase the infrared conditions are those at the horizon,
Eqs.~\eqref{eq:reg} and \eqref{eq:greg}, and we set $z_H=1$. We use the same
operators as in the confined phase. The charged black hole has a scaling
symmetry, so the condensates depend on $\muB$ and $T$ only through $\muB/T$.

\begin{figure}[h]
\centering
\subfloat[$\psi_i(z)$\label{fig:cbh_y}]{\includegraphics[width=0.49\linewidth]{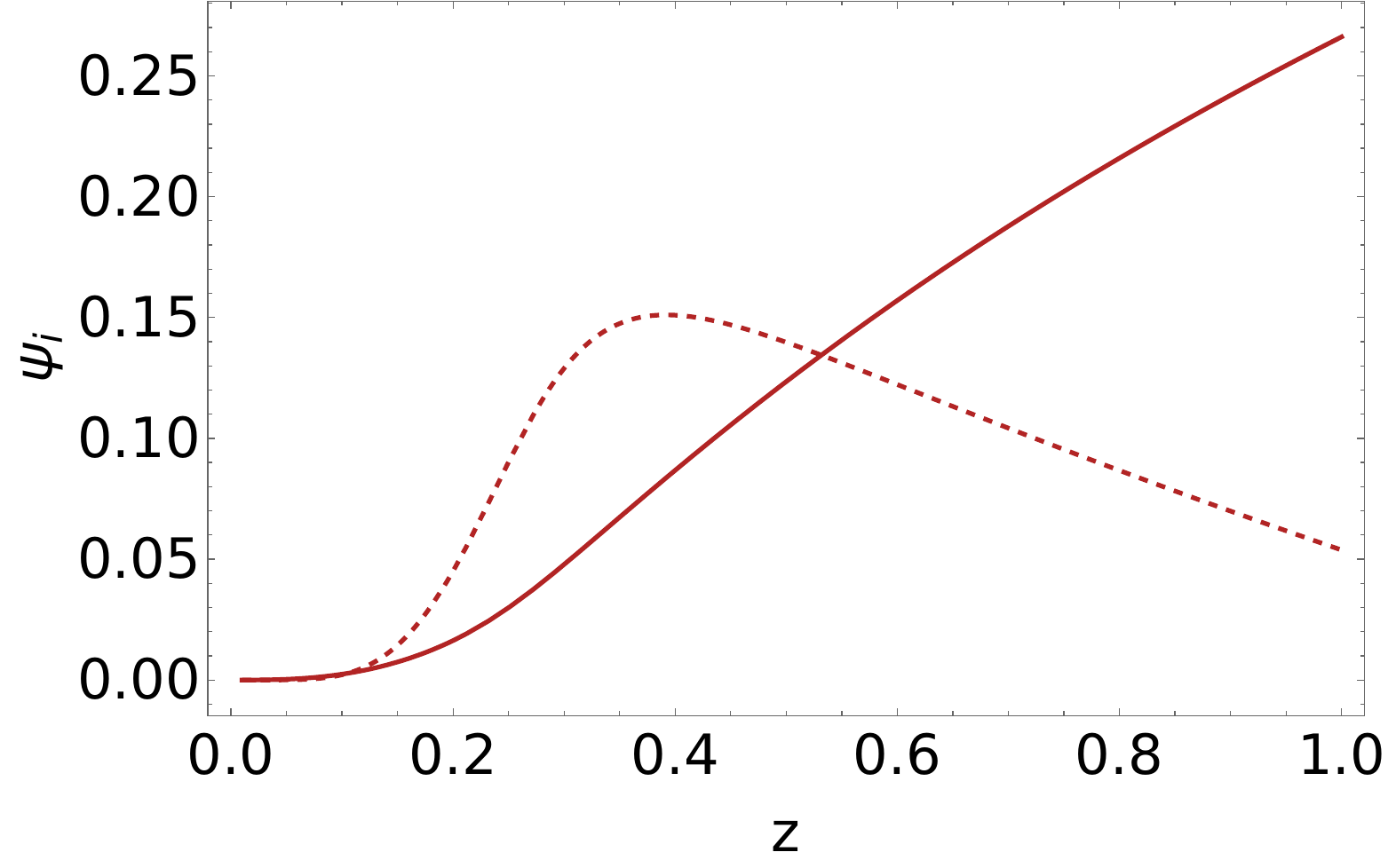}}\hfill
\subfloat[$\phi'(z)$\label{fig:cbh_fp}]{\includegraphics[width=0.49\linewidth]{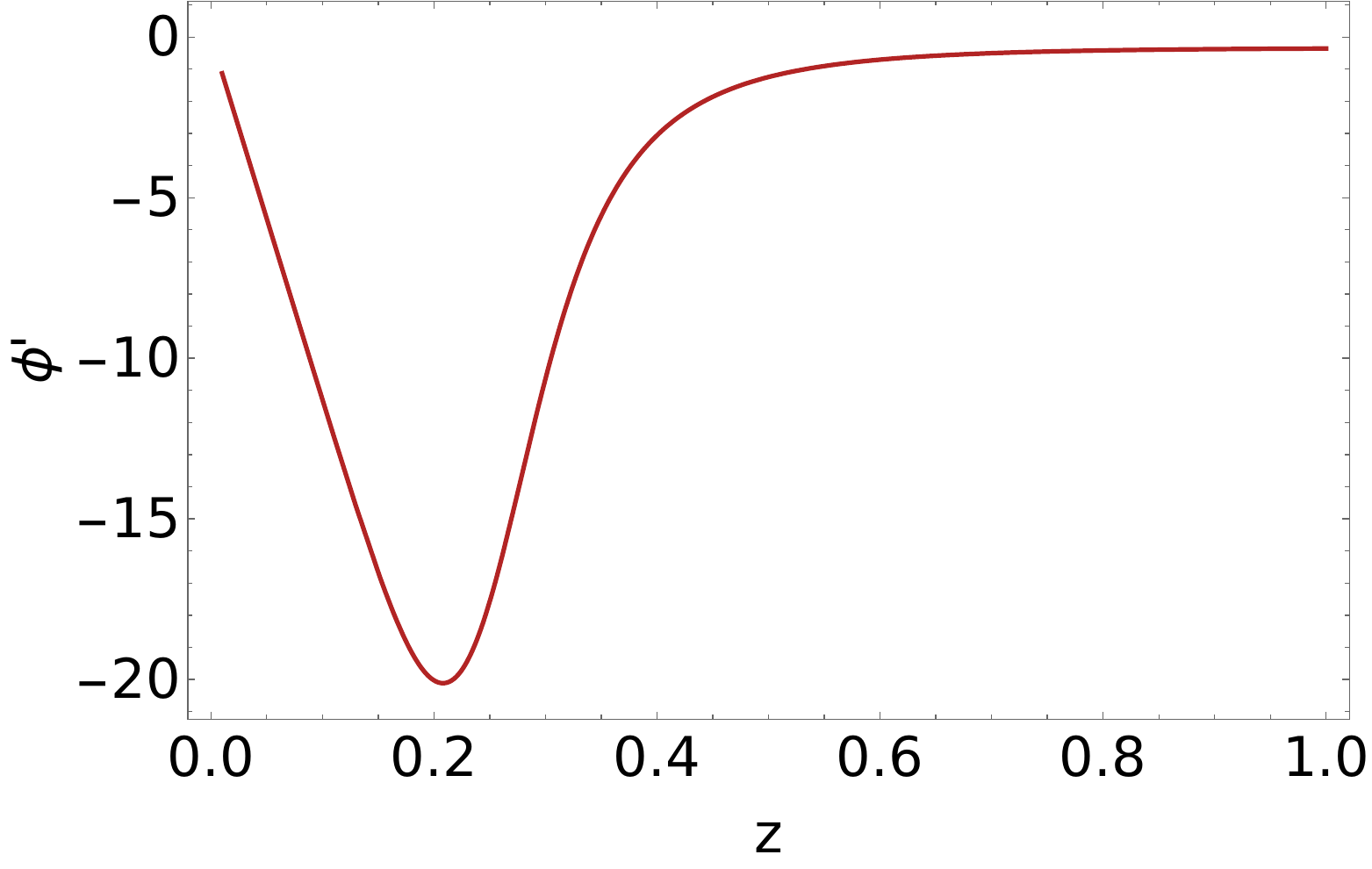}}
\caption{Charged black hole solution from the boundary to the horizon $z_H=1$ for
$q_1=2$, $m_1^2=-3$ (solid) and $q_2=6$, $m_2^2=10$ (dashed), at
$\muB/T\simeq260$ and $\zeta=0.77$.}
\label{fig:cbh_profiles}
\end{figure}

Figure~\ref{fig:cbh_profiles} shows the solution for the diquark-like and
dibaryon-like operators at $\muB/T\simeq260$, where both have condensed. The
profiles are smooth and nodeless, but they appear different from each other. The
diquark grows all the way to the horizon, while the dibaryon peaks at
$z\approx0.4\,z_H$ and decreases towards the horizon. The electric field is
strongly peaked at $z\approx0.2\,z_H$, in the region where the dibaryon profile
rises. The two condensates thus occupy different radial regions, unlike in the
confined phase, where both grow towards the wall (Fig.~\ref{fig:sol}).

\begin{figure}[h]
\centering
\subfloat[$q_2=3$, $m_2^2=-0.5$\label{fig:cbh_cond3}]{\includegraphics[width=0.46\linewidth]{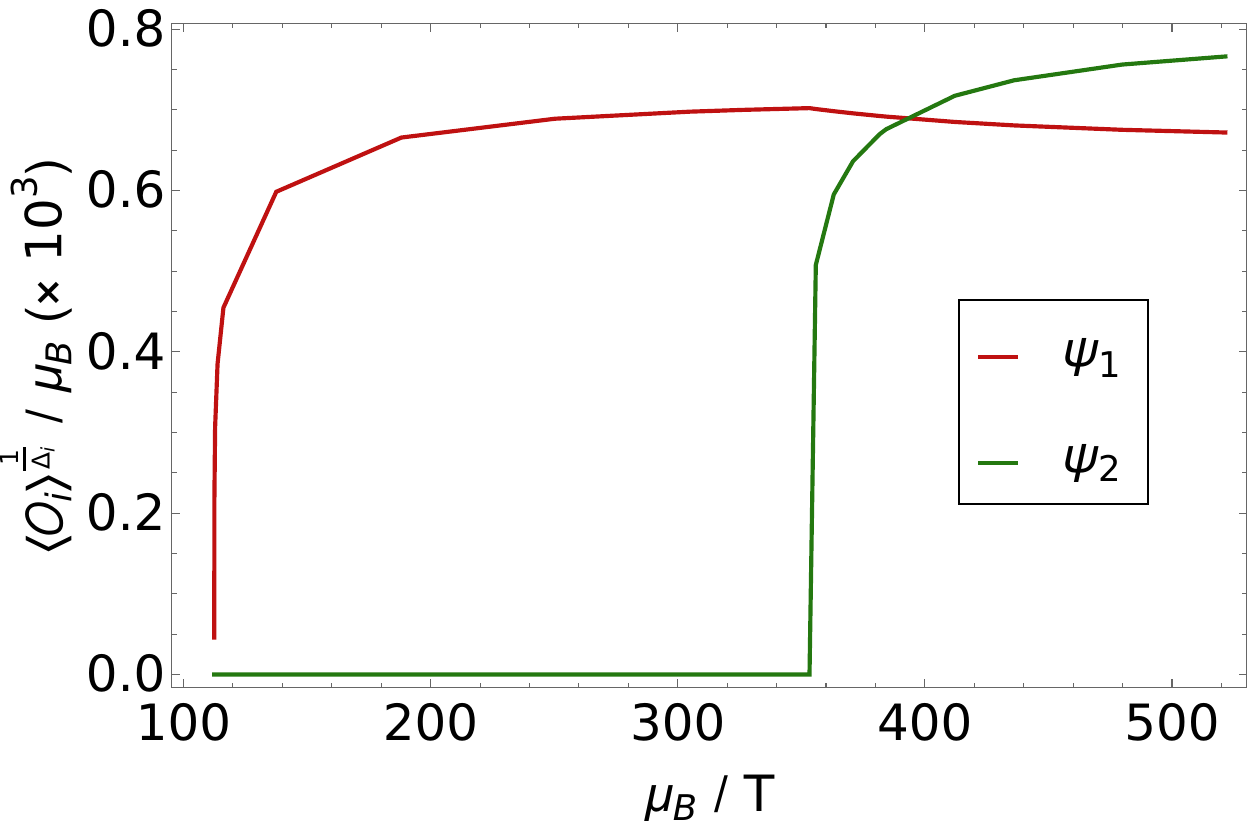}}\hfill
\subfloat[$q_2=6$, $m_2^2=10$\label{fig:cbh_cond6}]{\includegraphics[width=0.46\linewidth]{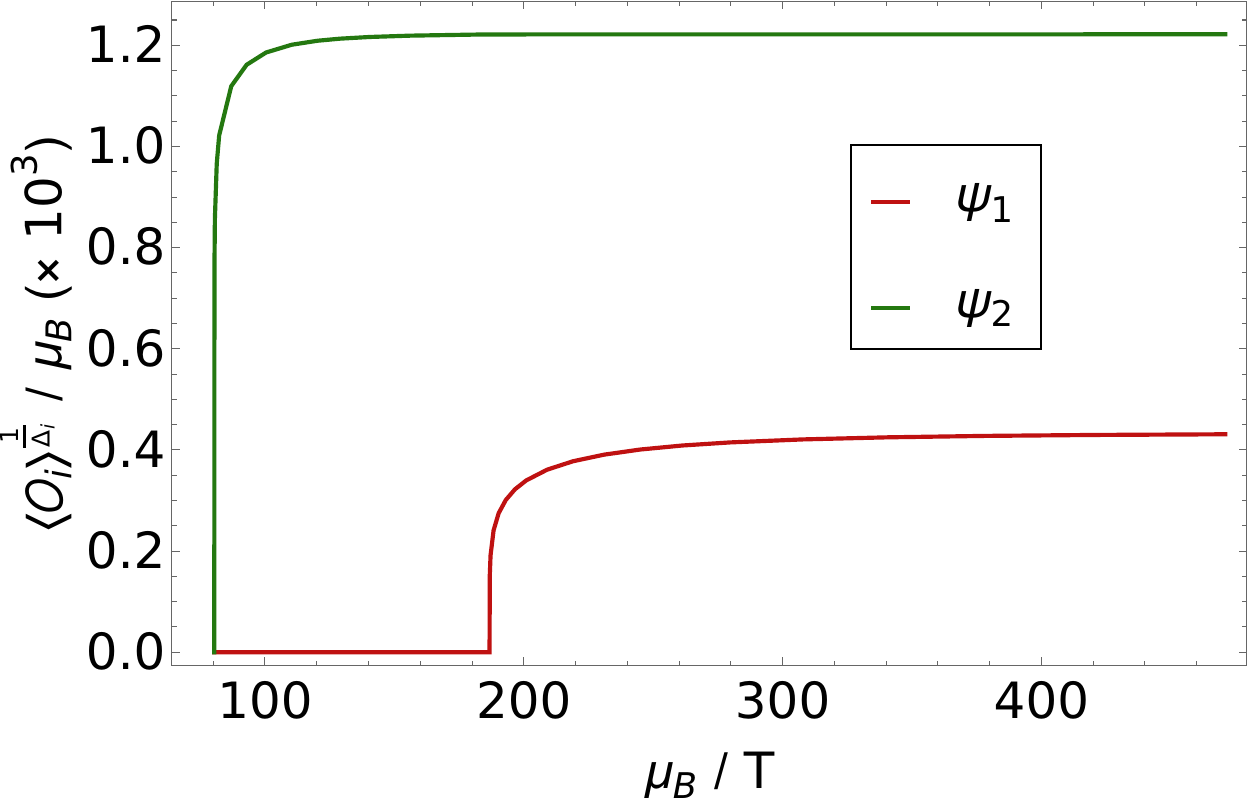}}
\caption{Condensates in the charged black hole as functions of $\muB/T$, for
$q_1=2$, $m_1^2=-3$ and $\zeta=0.77$.}
\label{fig:cbh_mubyT}
\end{figure}

\begin{figure}[h]
\centering
\subfloat[$q_2=3$\label{fig:cbh_pq3}]{\includegraphics[width=0.49\linewidth]{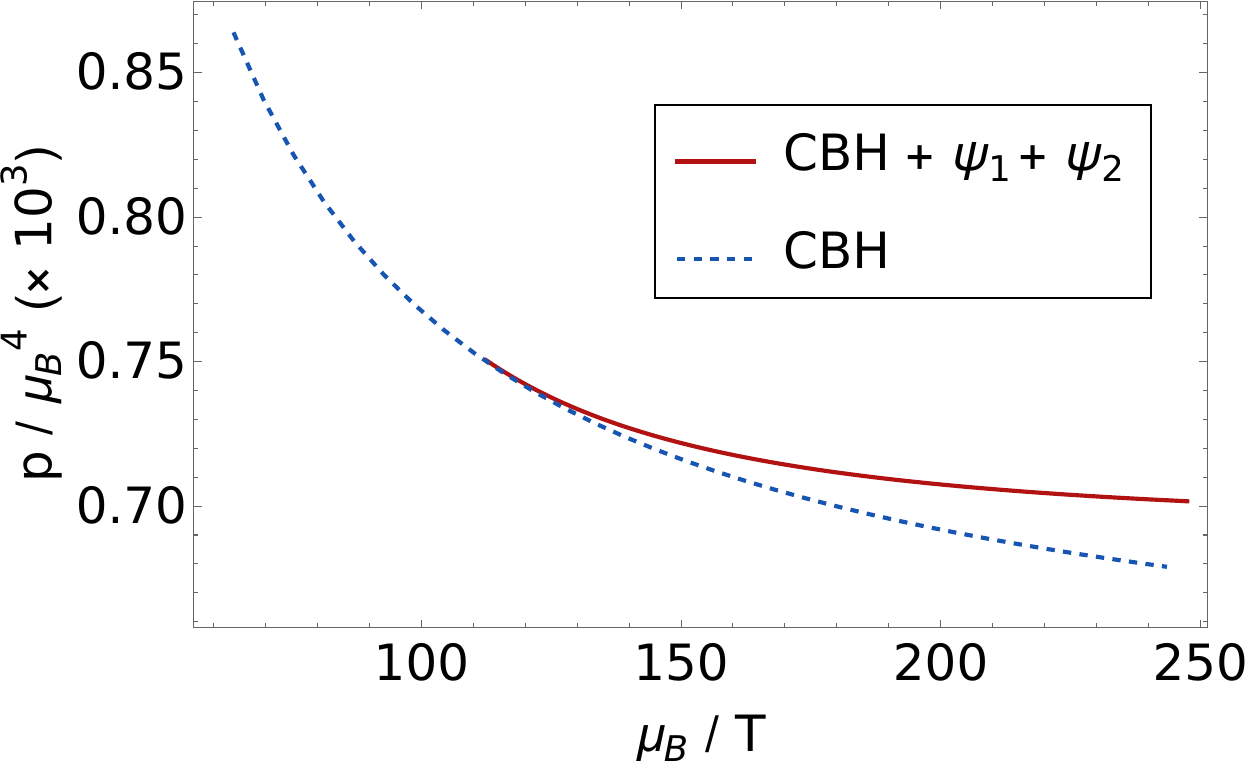}}\hfill
\subfloat[$q_2=6$\label{fig:cbh_pq6}]{\includegraphics[width=0.49\linewidth]{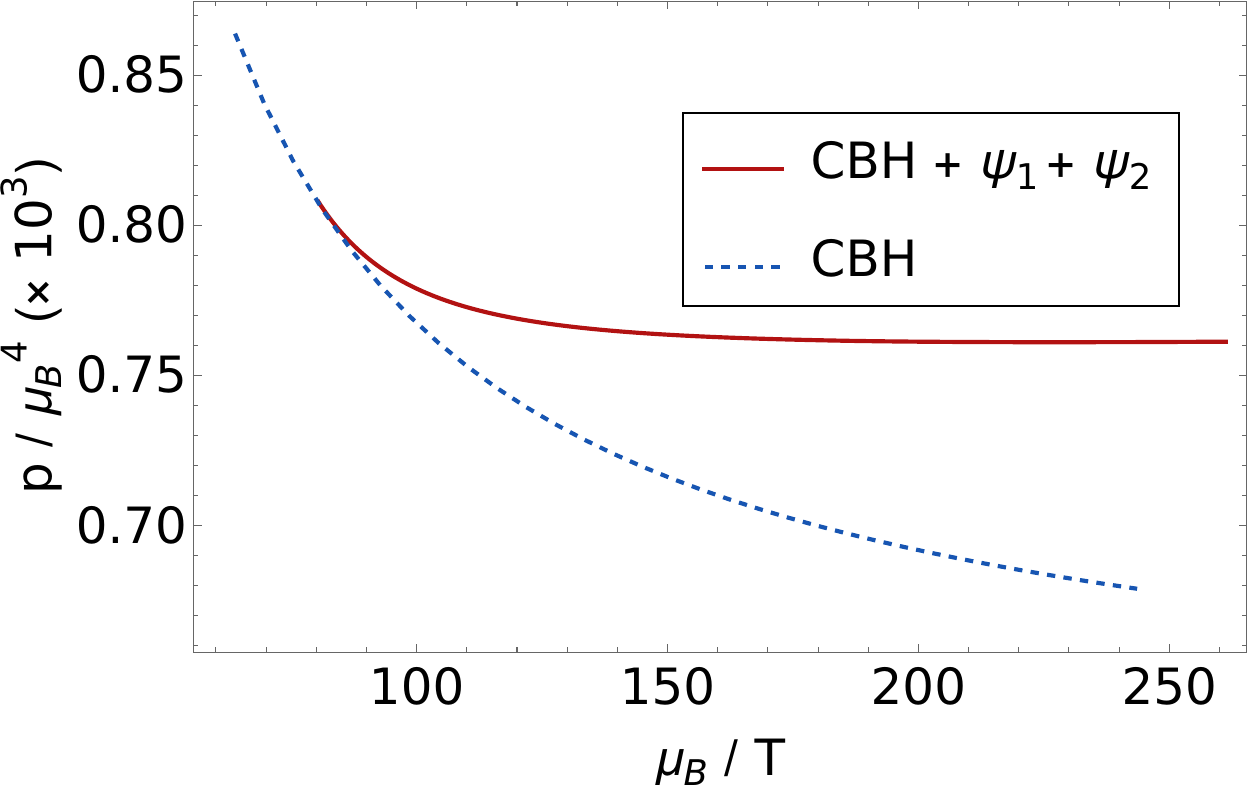}}
\caption{Pressure of the charged black hole with and without condensates as a function of $\muB/T$, for $q_1=2$, $m_1^2=-3$, the operators of Fig.~\ref{fig:cbh_mubyT} and $\zeta=0.77$.}
\label{fig:cbh_pressure}
\end{figure}

Figure~\ref{fig:cbh_mubyT} shows $\vev{\Op_i}^{1/\Delta_i}/\muB$ as functions of
$\muB/T$. The order of the onsets is the same as in the confined phase. For
$q_2=3$, $\Op_1$ condenses at $\muB/T\approx112$ and $\Op_2$ joins it at
$\muB/T\approx355$; the operator with the larger charge and mass condenses
second, as in Phase~B of Ref.~\cite{Cai:2013wma}. For $q_2=6$, the dibaryon
condenses first, at $\muB/T\approx80$, and the diquark joins it at
$\muB/T\approx188$; here the operator with the larger charge and mass condenses
first, as in their Phase~A. In both cases the two condensates persist once both
are present. As in the confined phase, the condensed solutions have the higher
pressure beyond each onset, and the onsets are continuous
(Fig.~\ref{fig:cbh_pressure}). In the phase diagrams of Sec.~\ref{sec:pd} the
first onset of each pair appears as a straight line through the origin. The
second lies inside the region where the confined phase is preferred. Therefore, it does not appear in the phase diagram.

\begin{figure}[h]
\centering
\subfloat[$q_2=3$\label{fig:cbh_m2a}]{\includegraphics[width=0.49\linewidth]{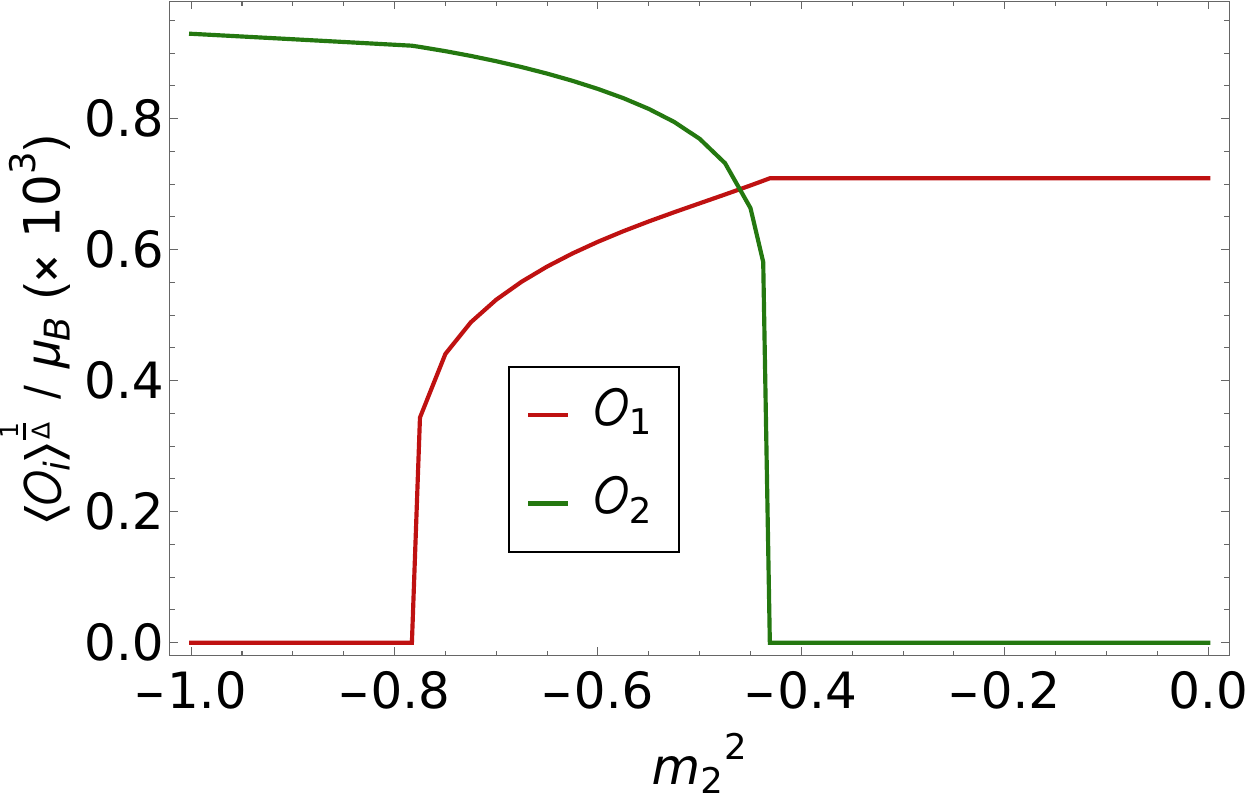}}\hfill
\subfloat[$q_2=6$\label{fig:cbh_m2b}]{\includegraphics[width=0.49\linewidth]{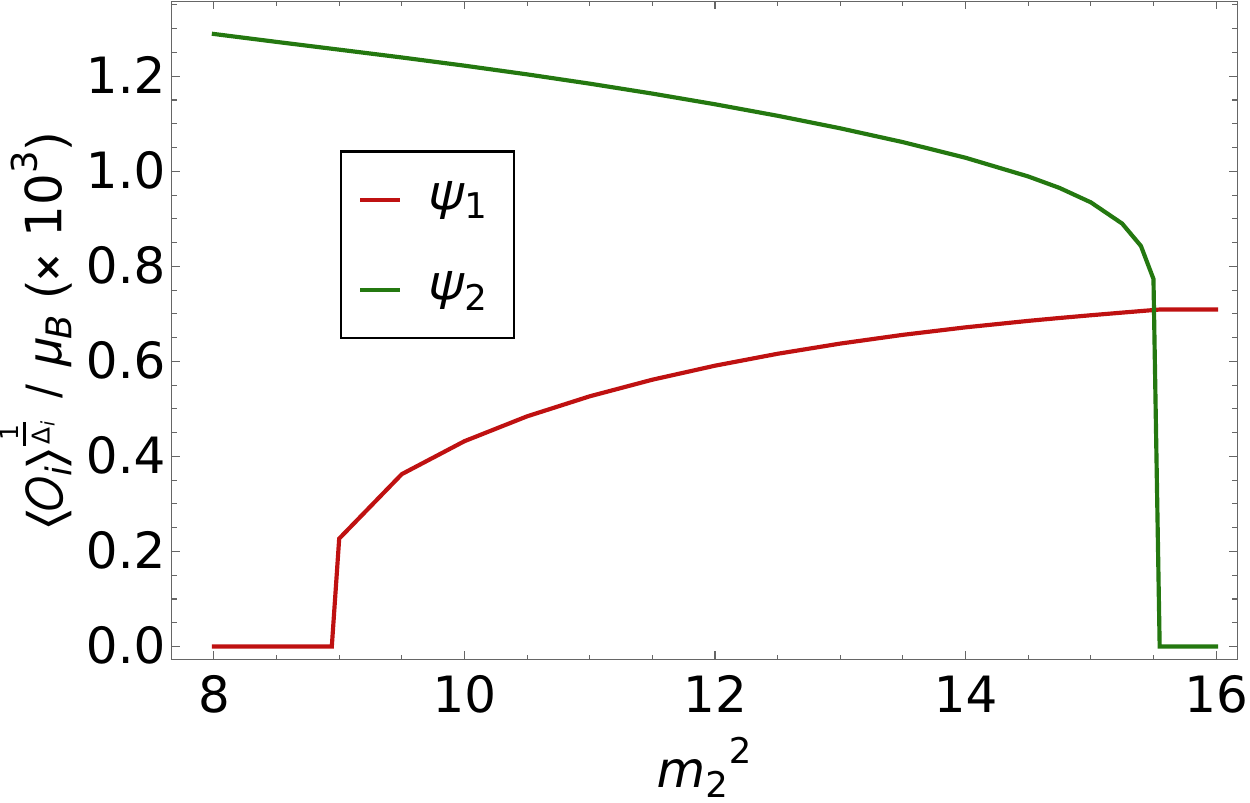}}
\caption{Condensates in the charged black hole at $\muB/T=540$ as functions of
$m_2^2$, for $q_1=2$, $m_1^2=-3$ and $\zeta=0.77$.}
\label{fig:cbh_window}
\end{figure}

At $\muB/T=540$ both condensates are nonzero, for $q_2=3$ as well as for $q_2=6$
(Fig.~\ref{fig:cbh_mubyT}). We fix $\muB/T$ at this value and vary $m_2^2$, as we
did at fixed $\mu$ in the confined phase. Figure~\ref{fig:cbh_window} shows that
the structure is the same as there. Only $\Op_1$ condenses at large $m_2^2$, only
$\Op_2$ at small $m_2^2$, and there is a coexistence window in between,
approximately $-0.78\lesssim m_2^2\lesssim-0.43$ ($3.79\lesssim\Delta_2\lesssim3.89$)
for $q_2=3$ and $9.0\lesssim m_2^2\lesssim15.6$ ($5.6\lesssim\Delta_2\lesssim6.4$)
for $q_2=6$. Inside the window the $\Op_1$ condensate decreases and the $\Op_2$
condensate increases as $m_2^2$ is lowered. The windows are wider than in the
confined phase, about $0.10$ against $0.07$ in $\Delta_2$ for $q_2=3$ and $0.8$
against $0.5$ for $q_2=6$. Since, the control parameter is $\muB/T$ here and $\mu$
there, this comparison is only indicative.

\section{Phase diagrams}\label{sec:pd}

\begin{figure}[h]
\centering
\subfloat[$q_2=3$, $m_2^2=-0.5$, $T=15$~MeV\label{fig:pt_q3}]{\includegraphics[width=0.45\linewidth]{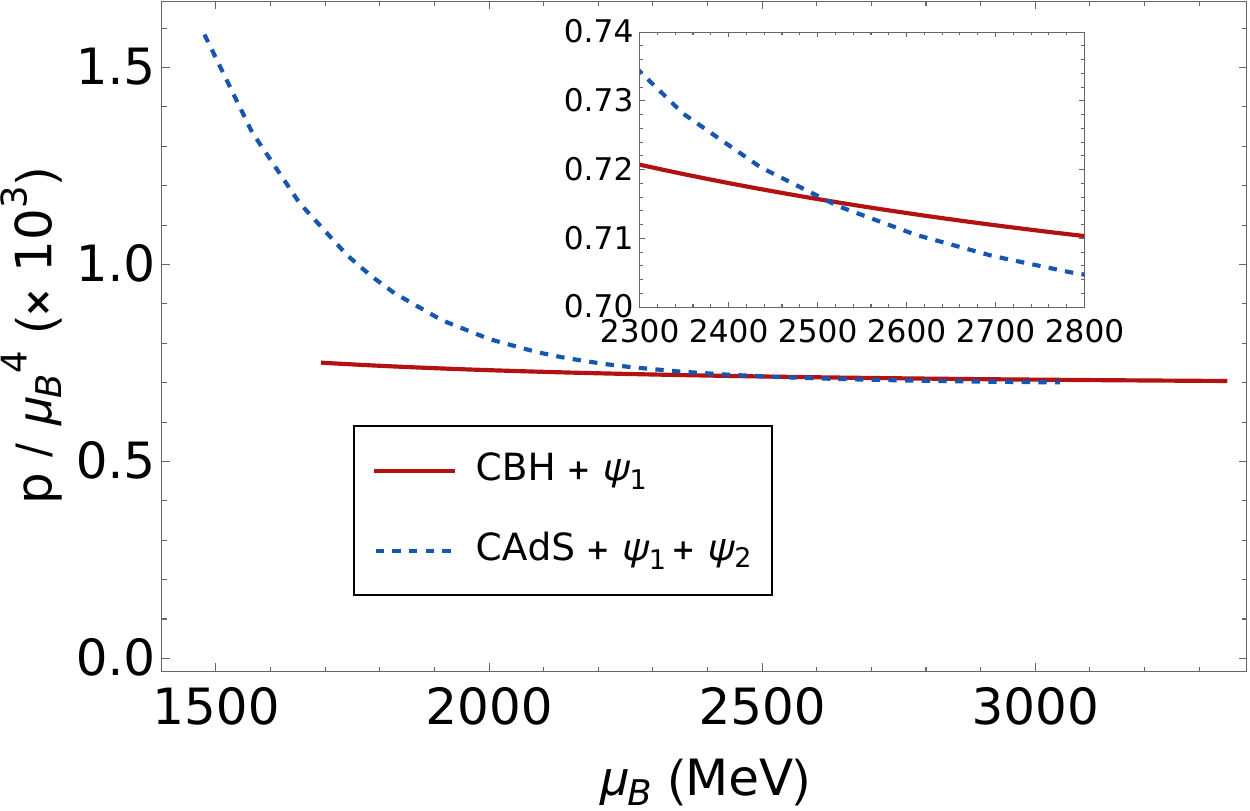}}\hfill
\subfloat[$q_2=6$, $m_2^2=10$, $T=28.5$~MeV\label{fig:pt_q6}]{\includegraphics[width=0.45\linewidth]{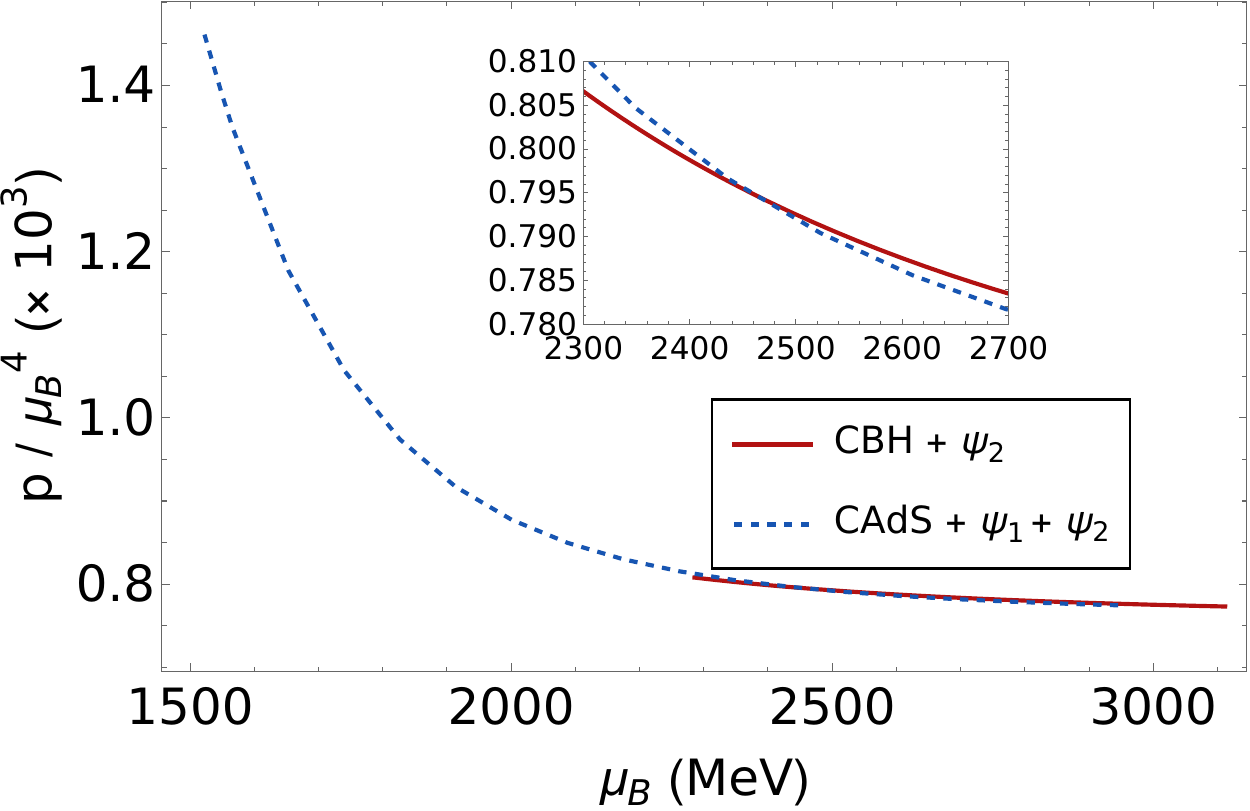}}
\caption{Pressure of the confined phase with both condensates (dashed) and of the
charged black hole with one condensate (solid) as functions of $\muB$, for
$q_1=2$, $m_1^2=-3$ and $\zeta=0.77$. The insets enlarge the region where the
two cross. The labels $\psi_1$, $\psi_2$ refer to $\Op_1$, $\Op_2$.}
\label{fig:pt}
\end{figure}

We compare the pressure of CBH and CAdS with condensates as a function of $\mu_B$ at fixed temperature. Figure~\ref{fig:pt} shows how the confined--deconfined boundary is obtained at
low temperature. For $q_2=3$ at $T=15$~MeV, the black hole with the $\Op_1$
condensate exists for $\muB\gtrsim1.69$~GeV, i.e.\ above its onset
$\muB/T\simeq112.5$. Its pressure lies below that of the confined phase with
both condensates up to $\muB\simeq2.51$~GeV and above it beyond
[Fig.~\ref{fig:pt_q3}]. For $q_2=6$ at $T=28.5$~MeV, the black hole with the
dibaryon condensate exists for $\muB\gtrsim2.29$~GeV ($\muB/T\simeq80.5$), and
the two pressures cross at $\muB\simeq2.47$~GeV [Fig.~\ref{fig:pt_q6}].

In both cases the curves cross at a finite angle. At the crossing, $p/\muB^4$
decreases about $2.8$ times faster on the confined branch for $q_2=3$, and about
$1.2$ times faster for $q_2=6$. The slope of the pressure therefore jumps, and
the transition is of first order, though only weakly so for $q_2=6$.

Now, we construct the phase diagrams in the $\muB$--$T$ plane by comparing, at each
$\muB$ and $T$, the pressures of all available solutions. CAdS with and without
condensates and the charged black hole with and without condensates. With
$\phi(z_0)=0$, the pressure of the horizonless solutions does not depend on
temperature, so condensation onsets in the confined phase are vertical blue lines. The blue dashed line represents the onset of $\Op_1$ and blue dotted-dashed line represents $\Op_2$. In the deconfined phase, they are straight lines through the origin, $T/\muB=\text{const}$ (green). Dashed lines refer to $\Op_1$ and dot-dashed lines to $\Op_2$. The red curves separate the confined and deconfined
phases; the solid part is the transition without condensates and the dashed
part the transition from the confined phase with condensates.

\begin{figure}[h]
\centering
\subfloat[$q_2=3$, $m_2^2=-0.5$\label{fig:pd_q3}]{\includegraphics[width=0.45\linewidth]{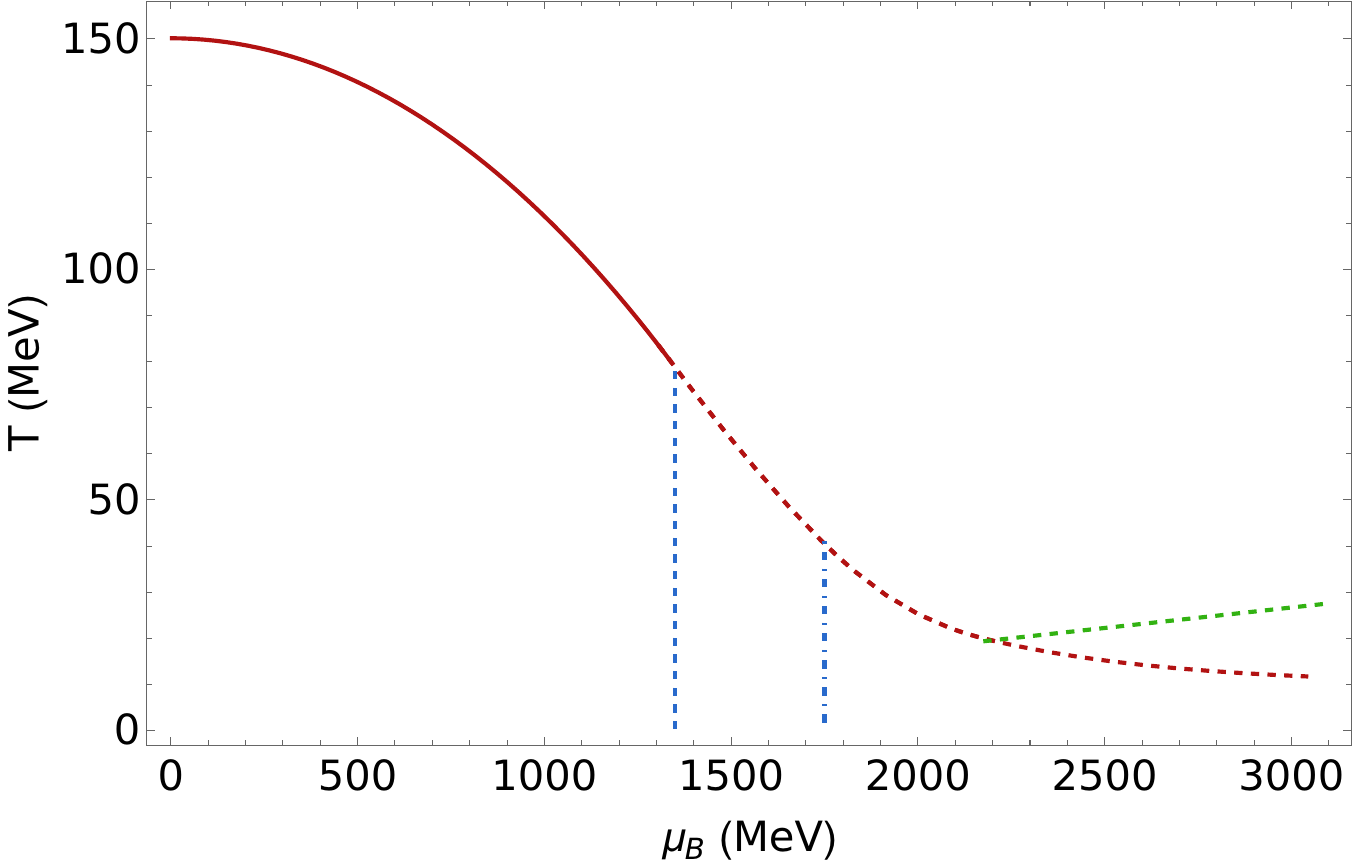}}\hfill
\subfloat[$q_2=6$, $m_2^2=10$\label{fig:pd_q6}]{\includegraphics[width=0.45\linewidth]{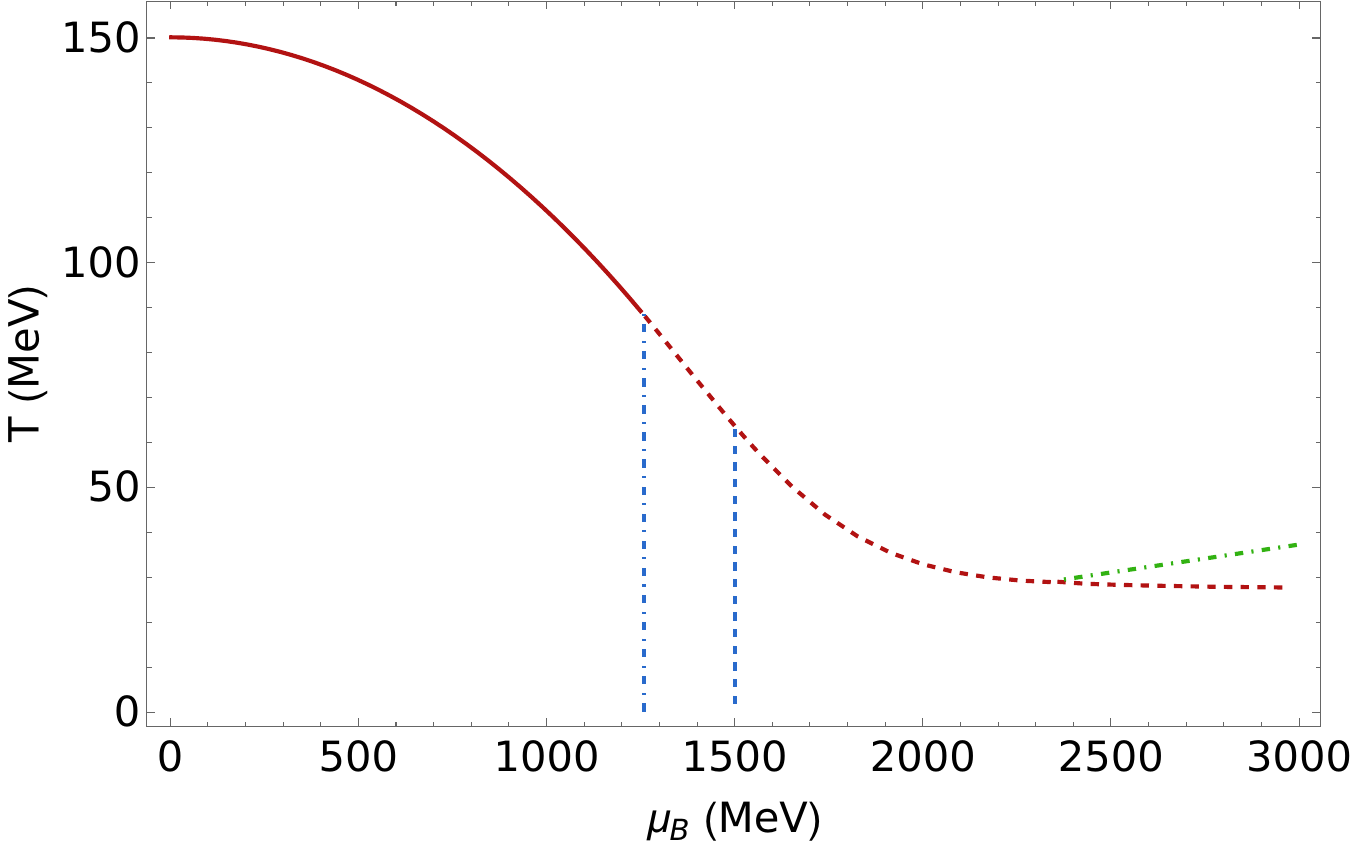}}
\caption{Phase diagrams in the $\muB$--$T$ plane for $q_1=2$, $m_1^2=-3$ and
$\zeta=0.77$. Red solid: transition between CAdS and the charged black hole
without condensates. Red dashed: first-order transition between the confined
phases with condensates and the deconfined phases. Blue: condensation onsets in
the confined phase; green: onsets in the deconfined phase. Dashed lines refer
to $\Op_1$ and dot-dashed lines to $\Op_2$.}
\label{fig:pd}
\end{figure}

For $q_2=3$ [Fig.~\ref{fig:pd_q3}], the confinement transition without
condensates falls from $T\simeq150$~MeV at $\muB=0$ to $T\simeq80$~MeV at
$\muB\simeq1.35$~GeV. There $\Op_1$ condenses in the confined phase, and $\Op_2$
joins it at $\muB\simeq1.75$~GeV . Beyond the first onset, the confined phase is
bounded from above by a first-order transition to the deconfined phase. Up to
$\muB\simeq2.18$~GeV the deconfined phase across this line is the black hole
without condensates. At $\muB\simeq2.18$~GeV and $T\simeq20$~MeV the line meets
the onset of $\Op_1$ in the black hole, $T\simeq\muB/112.5$. Beyond this point
the transition is to the black hole with the $\Op_1$ condensate, and the
transition temperature decreases only slowly, reaching $T\simeq12$~MeV at
$\muB\simeq3$~GeV. The onset of $\Op_2$ in the black hole,
$T\simeq\muB/355$, lies entirely below this line, so $\Op_2$ never condenses in
the deconfined phase.

For the dibaryon-like operator [Fig.~\ref{fig:pd_q6}] the order of the onsets
is reversed. The confinement transition without condensates reaches
$T\simeq88$~MeV at $\muB\simeq1.26$~GeV, where the dibaryon condenses in the
confined phase, and the diquark joins it at $\muB\simeq1.5$~GeV. The boundary
of the confined phase falls to $T\simeq29$~MeV at $\muB\simeq2.37$~GeV, where
it meets the onset of the dibaryon in the black hole, $T\simeq\muB/80.5$, and
then stays nearly flat at $T\simeq28$~MeV up to the largest $\muB$ computed.
The onset of the diquark in the black hole, $T\simeq\muB/188$, lies below the
transition line, so in the deconfined phase only the dibaryon condenses.

The two phase diagrams thus have the same structure. At low temperature the
confined phase with both condensates is preferred throughout the range of
$\muB$ we computed. Above it lies a narrow band of deconfined matter with a
single condensate, that of the operator with the lower threshold in $\muB/T$:
$\Op_1$ for $q_2=3$ and the dibaryon for $q_2=6$. Above that is the black hole
without condensates. The difference between the two cases is quantitative. At
$\muB\simeq3$~GeV the confined phase survives up to $T\simeq28$~MeV with the
dibaryon-like operator, but only up to $T\simeq12$~MeV with the $q_2=3$
operator. This reflects the larger pressure gain produced by the dibaryon
condensate in the confined phase, about a factor of three over CAdS at
$\muB\approx2.5$~GeV against about two for $q_2=3$
(Figs.~\ref{fig:q3_p} and \ref{fig:q6_p}).

\section{Discussion}\label{sec:discussion}

We have studied two condensing operators in a five-dimensional hardwall model
of dense matter, in the confined (charged AdS) and deconfined (charged black hole)
geometries. The two bulk scalars carry independent charges and dimensions and
interact only through the gauge field and the metric. With the infrared
amplitudes fixed by maximizing the pressure, the two operators condense together
only within a narrow window of operator dimensions. At fixed $q_1=2$,
$\Delta_1=3$ and $\mu=2.2$ ($\muB\simeq1.91$~GeV), the window has a width of
$0.07$, $0.21$ and $0.50$ in $\Delta_2$ for $q_2=3$, $4$ and $6$, and it closes
at the degenerate point $(q_2,\Delta_2)=(2,3)$. In the $(q_2,\Delta_2)$ plane
this coexistence strip (Phase~III) separates the region in which only $\Op_1$
condenses (Phase~I) from the region in which only $\Op_2$ condenses (Phase~II).
At both edges of the strip the vanishing condensate goes to zero continuously. The narrowness of
the windows suggests that the electrostatic repulsion dominates.

As functions of the chemical potential, both operators condense through
continuous transitions. For the interpolating charge $q_2=3$ the diquark-like
operator condenses first, at $\muB\simeq1.35$~GeV, and the second operator joins
it at $\muB\simeq1.75$~GeV. For the dibaryon-like operator the order is
reversed, the dibaryon condenses at $\muB\simeq1.26$~GeV and the diquark joins
it at $\muB\simeq1.5$~GeV. Once both are present they persist, and they raise
the pressure of the confined phase well above that of CAdS without condensates.
In the deconfined phase the same window structure appears, with somewhat wider
windows, and the dibaryon and diquark profiles occupy different radial regions
of the black hole. 

At low temperature, the confined phase with both condensates is preferred
throughout the range of $\muB$ we computed. It is bounded from above by a
first-order transition to a narrow band of deconfined matter carrying a single
condensate, $\Op_1$ for $q_2=3$ and the dibaryon for $q_2=6$. The confined phase
survives to higher temperature with the dibaryon-like operator, up to about
$28$~MeV against $12$~MeV at $\muB\simeq3$~GeV, which reflects the larger
pressure gain produced by the dibaryon condensate.

For the physically relevant pair, the diquark ($q=2$) and the dibaryon
($q=2N_c=6$), the results point to a sequence in which the dibaryon condenses
first and the diquark joins it at higher density, provided the dimension of the
dibaryon operator falls inside the window. The perturbative estimate $\Delta=9$
lies far outside it. Coexistence therefore requires a dibaryon operator that is
strongly bound in the medium, with a dimension close to $6$. It is suggestive
that this is also the range, $\Delta\gtrsim5$--$6$, that we found in
~\cite{Singh:2025ogp} to be compatible with an onset of dibaryon condensation
above the nucleon mass, although that analysis used phenomenological infrared
conditions and the comparison is only qualitative. Within the model, whether
dense matter supports both condensates thus reduces to a question about one
number, the in-medium dimension of the dibaryon operator.

We have not included direct couplings between the scalars, but several are
allowed by gauge invariance: self-interactions $|\psi_i|^4$, a scalar--scalar
coupling $|\psi_1|^2|\psi_2|^2$ and, when $q_2=nq_1$, terms
$\psi_2^*\psi_1^n+\text{c.c.}$, which are cubic for $q_2=4$ and quartic for the
dibaryon, $q_2=6$. 
The scalar--scalar
coupling vanishes whenever one of the scalars vanishes, so the single-condensate
phases do not depend on it. The pressure of the coexistence phase, in
contrast, changes by
$\delta p\propto\int dz\,\sqrt{-g}\,\psi_1^2\psi_2^2$. The sign of the coupling can narrow or widen the coexistence window size. Because the windows are
narrow, a modest change in the strength of coupling might change them qualitatively. From the QCD point of view, other operators will matter
as well. A bulk field dual to the chiral condensate couples to both scalars
through terms such as $|X|^2|\psi_i|^2$. Distinguishing neutron and proton
pairing requires an isospin chemical potential~\cite{Kovensky:2021ddl}.
Spin-triplet pairing has a vector order parameter, as in holographic $p$-wave
superconductors~\cite{Gubser:2008wv, Cai:2015cya}, which brings new couplings to the gauge
field beyond those fixed by gauge invariance.
Several extensions suggest themselves. The most immediate is to repeat the
analysis with the phenomenological infrared conditions of
Refs.~\cite{Singh:2024amm,Singh:2025ogp}, for which the density obtained from the
pressure agrees with the charge of the solution, and which give phase diagrams
that can be compared directly with effective field theories in their regime of
validity. Computing the phase diagrams with the direct couplings discussed above
switched on is a natural next step. The fluctuations around the coexistence
phase would show whether coexistence leaves an imprint on transport. Finally, the
large increase in pressure produced by the dibaryon condensate suggests that its
effect on the equation of state, and hence on the structure of neutron stars,
deserves a closer look once it is computed with phenomenological boundary
conditions.

\section*{Data availability}
The data that support the findings of this article are openly available~\cite{Singh:2026zenodo}.

\begin{acknowledgments}
I thank K.P. Yogendran for comments on the manuscript. This work is
supported by the South African Research Chairs Initiative of the Department of
Science, Technology, and Innovation (DSTI) and the National Research Foundation
(NRF), grant SARC20110101-078554.
\end{acknowledgments}

\appendix

\section{Holographic renormalization}\label{app:renorm}

We follow the scheme of~\cite{Singh:2025ogp}, generalized to two scalars. Using the equations of motion, the bulk action~\eqref{eq:action} evaluates on shell to a total derivative~\cite{Sonner:2010yx},
\begin{equation}\label{eq:Sbulk}
S_{\text{bulk}}=\frac{1}{\kappa^2}\int d^5x\,\frac{d}{dz}\left(\frac{1}{z}\sqrt{-g}\,g^{zz}\right).
\end{equation}
At the wall this depends on $g(z_0)$, $h(z_0)$ and the gauge field there, but not on $\psi_i(z_0)$, which is what permits the maximum-pressure prescription for both scalars. The action diverges at the boundary and is renormalized as~\cite{deHaro:2000vlm}
\begin{equation}
S_{\text{ren}}=S_{\text{bulk}}+S_{\text{GH}}+S_{\text{ct},1}+S_{\text{ct},2},
\end{equation}
with
\begin{align}
S_{\text{GH}}&=-\frac{1}{\kappa^2}\int d^4x\sqrt{\gamma}\,K,\qquad
S_{\text{ct},1}=\frac{3}{\kappa^2L}\int d^4x\sqrt{\gamma},\nonumber\\
S_{\text{ct},2}&=\sum_{i=1}^2\frac{\lambda_i(4-\Delta_{i+})}{L}\int d^4x\sqrt{\gamma}\,\psi_i^2,
\end{align}
where $\gamma$ is the induced metric at $z=\epsilon$ and $K$ the trace of its extrinsic curvature. No Gibbons--Hawking term is added at the wall, consistent with its interpretation as an excision surface behind which the sources reside. $S_{\text{ct},1}$ cancels the divergences of the gravitational action and the Gibbons--Hawking term; the Maxwell term is finite. The scalar counterterm is a sum of two independent terms, each fixed by its own $\Delta_{i+}$. Since the scalars do not mix in the action, no counterterm proportional to $\psi_1\psi_2$ is generated.

The pressure is $p=-\Omega$, with $\Omega$ the renormalized on-shell free-energy density. The baryon density is given by $\rho_B=\partial p/\partial \muB$, the entropy density is $s=\partial p/\partial T$, and the energy density follows from the Euler relation $\varepsilon=-p+Ts+\muB\rho_B$. The boundary stress tensor cannot be used for this purpose, because part of the geometry has been neglected~\cite{Singh:2025ogp}.

\section{Numerical method and dependence on the ultraviolet cutoff}\label{app:cutoff}

We solve the equations with a shooting method. For given infrared amplitudes $(\psi_{10},\psi_{20})$ we integrate from the wall to $z=\epsilon$, adjusting the remaining infrared data until both scalar sources vanish, $h(\epsilon)=1$ and $\phi(\epsilon)=\mu$. The pressure is evaluated on a grid in the $(\psi_{10},\psi_{20})$ plane and maximized. The two-field problem is stiffer than the single-field one, especially at large $\Delta_2$, where keeping the non-normalizable mode at zero is delicate. The maximum-pressure solutions reported here lie well inside the curve on which $g_{tt}(z_0)\to0$.

The boundary conditions and the renormalized action are imposed at $z=\epsilon$, so every quantity carries a cutoff error, and since the ground state is selected by maximizing $p$ over the infrared data, an error in $p$ propagates into the location of the maximum. To quantify this we take $q_1=2$, $m_1^2=-3$, $q_2=6$, $m_2^2=10$, $g_0=6$, $z_0=1$, for which the ground state is the coexistence phase, and repeat the full maximization at several values of $\epsilon$ for $\mu=1.8$, $2.2$ and $2.6$. Table~\ref{tab:cutoff} lists the results.

\begin{table}[t]
\caption{Maximum-pressure solution as a function of the ultraviolet cutoff for $q_1=2$, $m_1^2=-3$, $q_2=6$, $m_2^2=10$, $g_0=6$, $z_0=1$. $\Omega=-p$ is the renormalized free-energy density. The ground state is the coexistence phase in all cases.}
\label{tab:cutoff}
\begin{ruledtabular}
\begin{tabular}{lcccc}
$\epsilon$ & $\psi_{10}$ & $\psi_{20}$ & $\Omega$ & $\vev{\Op_1}$\\
\hline
\multicolumn{5}{l}{$\mu=1.8$}\\
$10^{-1}$        & $0.4214$ & $0.1504$ & $-1.16716$ & $1.1129$\\
$5\times10^{-2}$ & $0.2894$ & $0.2411$ & $-1.15075$ & $0.8226$\\
$10^{-2}$        & $0.1927$ & $0.2719$ & $-1.14760$ & $0.5707$\\
$5\times10^{-3}$ & $0.1884$ & $0.2729$ & $-1.14753$ & $0.5589$\\
$10^{-4}$        & $0.1870$ & $0.2732$ & $-1.14751$ & $0.5548$\\
\multicolumn{5}{l}{$\mu=2.2$}\\
$10^{-1}$        & $0.4081$ & $0.3223$ & $-1.81250$ & $1.1748$\\
$5\times10^{-2}$ & $0.2906$ & $0.3624$ & $-1.75478$ & $0.9072$\\
$10^{-2}$        & $0.2085$ & $0.3788$ & $-1.73978$ & $0.6799$\\
$5\times10^{-3}$ & $0.2050$ & $0.3794$ & $-1.73935$ & $0.6693$\\
$10^{-4}$        & $0.2037$ & $0.3796$ & $-1.73920$ & $0.6657$\\
\multicolumn{5}{l}{$\mu=2.6$}\\
$10^{-1}$        & $0.5059$ & $0.2975$ & $-3.24702$ & $1.6340$\\
$5\times10^{-2}$ & $0.3609$ & $0.3743$ & $-3.06573$ & $1.2983$\\
$10^{-2}$        & $0.2353$ & $0.4057$ & $-3.01823$ & $0.9113$\\
$5\times10^{-3}$ & $0.2290$ & $0.4068$ & $-3.01686$ & $0.8897$\\
$10^{-4}$        & $0.2268$ & $0.4072$ & $-3.01641$ & $0.8823$\\
\end{tabular}
\end{ruledtabular}
\end{table}

All quantities approach their limits as $\epsilon^2$, as expected from the near-boundary expansion. Halving $\epsilon$ from $10^{-2}$ to $5\times10^{-3}$ reduces the error in $\Omega$ by a factor between $4.0$ and $4.3$. For the amplitudes the factor lies between $3.6$ and $6.6$, the larger spread reflecting the flatness of the pressure along one direction in the plane, which amplifies residual noise in the location of the maximum. Writing $X(\epsilon)-X(0)\simeq c_X\epsilon^2$, we find at $\mu=2.2$ $c_{\psi_{10}}\simeq54$, $c_{\psi_{20}}\simeq-7.9$ and $c_\Omega\simeq-5.7$. The coefficient $c_\Omega$ grows quickly with $\mu$, by a factor of about twenty between $\mu=1.8$ and $2.6$.

The observables are not equally sensitive. The pressure is the most robust: the solution is a stationary point, so the error in $\Omega$ is quadratic in the displacement of the maximum, and at $\epsilon=10^{-2}$ it is below $0.1\%$ at all three values of $\mu$. The amplitude $\psi_{10}$ is the most sensitive; at $\epsilon=10^{-1}$ it exceeds its limit by more than a factor of two, and at $\epsilon=10^{-2}$ it is off by $2$--$4\%$, as is $\vev{\Op_1}$. The error in $\psi_{20}$ at $\epsilon=10^{-2}$ is below $0.5\%$.

\bibliography{refs}

\end{document}